%% file: main.tex
\documentclass[5p]{elsarticle}

\input{math_commands.tex}

\usepackage{multirow}
\usepackage[utf8]{inputenc} 
\usepackage[T1]{fontenc}    
\usepackage{amsmath}
\usepackage{hyperref}       
\usepackage{url}            
\usepackage{booktabs}       
\usepackage{amsfonts}       
\usepackage{nicefrac}       
\usepackage{microtype}      
\usepackage{lipsum}		
\usepackage[dvipsnames]{xcolor}
\usepackage{graphicx}
\usepackage{doi}
\usepackage{caption}
\usepackage{subcaption}
\usepackage{enumitem}
\usepackage{booktabs}

\usepackage[normalem]{ulem}
\useunder{\uline}{\ul}{}

\newcommand{\xat}{\textcolor{Blue}{\mathbf{x}^{(a)}_t}}
\newcommand{\xatT}{\textcolor{Blue}{\mathbf{x}^{(a)}_{t:T}}}
\newcommand{\xa}{\textcolor{Blue}{\mathbf{x}^{(a)}}}

\newcommand{\xvt}{\textcolor{Red}{\mathbf{x}^{(v)}_t}}
\newcommand{\xvtT}{\textcolor{Red}{\mathbf{x}^{(v)}_{t:T}}}
\newcommand{\xv}{\textcolor{Red}{\mathbf{x}^{(v)}}}

\newcommand{\zvt}{\textcolor{Red}{\mathbf{z}^{(v)}_t}}
\newcommand{\zv}{\textcolor{Red}{\mathbf{z}^{(v)}}}
\newcommand{\zvtt}{\textcolor{Red}{\mathbf{z}^{(v)}_{1:t-1}}}

\newcommand{\zat}{\textcolor{Blue}{\mathbf{z}^{(a)}_t}}
\newcommand{\za}{\textcolor{Blue}{\mathbf{z}^{(a)}}}
\newcommand{\zatt}{\textcolor{Blue}{\mathbf{z}^{(a)}_{1:t-1}}}

\newcommand{\zavt}{\textcolor{OliveGreen}{\mathbf{z}^{(av)}_t}}
\newcommand{\zav}{\textcolor{OliveGreen}{\mathbf{z}^{(av)}}}
\newcommand{\zavtt}{\textcolor{OliveGreen}{\mathbf{z}^{(av)}_{1:t-1}}}

\newcommand{\x}{\mathbf{x}}
\newcommand{\z}{\mathbf{z}}
\newcommand{\w}{\textcolor{Green}{\mathbf{w}}}

\makeatletter
\def\ps@pprintTitle{%
 \let\@oddhead\@empty
 \let\@evenhead\@empty
 \def\@oddfoot{}%
 \let\@evenfoot\@oddfoot}
\makeatother

\hypersetup{
    colorlinks=true,
    linkcolor=blue,
    filecolor=magenta,      
    urlcolor=cyan
}

\biboptions{authoryear}

\begin{document}

\begin{frontmatter}
\title{A Multimodal Dynamical Variational Autoencoder for Audiovisual Speech Representation Learning\tnoteref{t1}}

\author[1]{Samir Sadok\corref{cor1}}
\ead{samir.sadok@centralesupelec.fr}
\author[1]{Simon Leglaive}
\author[2]{Laurent Girin}
\author[3]{Xavier Alameda-Pineda}
\author[1]{Renaud S\'eguier}
\address[1]{CentraleSupélec, IETR UMR CNRS 6164, France}
\address[2]{Univ.~Grenoble Alpes, CNRS, Grenoble-INP, GIPSA-lab, France}
\address[3]{Inria, Univ.~Grenoble Alpes, CNRS, LJK, France}
\cortext[cor1]{Corresponding author}
\tnotetext[t1]{This work was supported by Randstad corporate research chair, by ANR-3IA MIAI (ANR-19-P3IA-0003), by ANR-JCJC ML3RI (ANR-19-CE33-0008-01), and by H2020 SPRING (funded by EC under GA \#871245). }

\begin{abstract}
\label{abstract}
High-dimensional data such as natural images or speech signals exhibit some form of regularity, preventing their dimensions from varying independently. This suggests that there exists a lower dimensional latent representation from which the high-dimensional observed data were generated. Uncovering the hidden explanatory features of complex data is the goal of representation learning, and deep latent variable generative models have emerged as promising unsupervised approaches. In particular, the variational autoencoder (VAE) which is equipped with both a generative and an inference model allows for the analysis, transformation, and generation of various types of data. Over the past few years, the VAE has been extended to deal with data that are either multimodal \textit{or} dynamical (i.e., sequential).
In this paper, we present a multimodal \textit{and} dynamical VAE (MDVAE) applied to unsupervised audiovisual speech representation learning. The latent space is structured to dissociate the latent dynamical factors that are shared between the modalities from those that are specific to each modality. A static latent variable is also introduced to encode the information that is constant over time within an audiovisual speech sequence. The model is trained in an unsupervised manner on an audiovisual emotional speech dataset, in two stages. In the first stage, a vector quantized VAE (VQ-VAE) is learned independently for each modality, without temporal modeling. The second stage consists in learning the MDVAE model on the intermediate representation of the VQ-VAEs before quantization. The disentanglement between static versus dynamical and modality-specific versus modality-common information occurs during this second training stage.
Extensive experiments are conducted to investigate how audiovisual speech latent factors are encoded in the latent space of MDVAE. These experiments include manipulating audiovisual speech, audiovisual facial image denoising, and audiovisual speech emotion recognition. The results show that MDVAE effectively combines the audio and visual information in its latent space. They also show that the learned static representation of audiovisual speech can be used for emotion recognition with few labeled data, and with better accuracy compared with unimodal baselines and a state-of-the-art supervised model based on an audiovisual transformer architecture.
\end{abstract}


\end{frontmatter}


\section{Introduction and related work}
The world around us is represented by a multitude of different modalities \citep{lazarus1976multimodal}. A single event can be observed from different perspectives, and combining these different views can provide a complete understanding of what is happening. For instance, speech in human interactions is a multimodal process where the audio and visual modalities carry complementary verbal and non-verbal information.
By capturing the correlations between different modalities, we can reduce uncertainty and better understand a phenomenon \citep{bengio2013representation}. Combining complementary sources of information from heterogeneous modalities is a challenging task, for which machine and deep learning techniques have shown their efficiency. In particular, the flexibility and versatility of deep neural networks allow them to efficiently learn from heterogeneous data to solve a given task \citep{ramachandram2017deep, baltruvsaitis2018multimodal}.

The rapid development of artificial intelligence technology and hardware acceleration has led to a shift towards multimodal processing \citep{ramachandram2017deep}, which aims to enhance machine perception by integrating various data types. With the explosion of digital content and communication, audiovisual speech  processing has become increasingly important for a range of applications, such as speech recognition \citep{afouras2018deep, petridis2018end, hori2019end}, speaker identification \citep{roth2020ava}, and emotion recognition \citep{wu2014survey, noroozi2017audio, schoneveld2021leveraging}. 
However, in tasks such as emotion recognition, the limited availability of labeled data remains a significant challenge. As a result, researchers are investigating unsupervised or weakly supervised methods to learn effective audiovisual speech  representations. This is extremely promising in problem settings involving a large amount of unlabeled data but limited labeled data.

Deep generative models \citep{kingma2014auto, rezende2014stochastic, goodfellow2014generative} have recently become very successful for unsupervised learning of latent representations from high-dimensional and structured data such as images, audio, and text. Learning meaningful representations is essential not only for synthesizing data but also for data analysis and transformation. 
For a learned representation to be effective, it should capture high-level characteristics that are invariant to small and local changes in the input data, and it should be as disentangled as possible for explainability. Furthermore, hierarchical and disentangled generative models have demonstrated their efficacy to solve downstream learning tasks \citep{van2019disentangled, bengio2013representation}.
Variants of generative models have recently led to considerable progress in disentangled representation learning, particularly with the variational autoencoder (VAE) \citep{kingma2014auto, rezende2014stochastic}. 

The VAE considers that an observed high-dimensional data vector $\mathbf{x}$ is generated by a low-dimensional latent vector $\mathbf{z}$. The generative process is characterized by the joint distribution $p_\theta(\mathbf{x},\mathbf{z}) = p_\theta(\mathbf{x} | \mathbf{z}) p(\mathbf{z})$, where $p(\mathbf{z})$ is the prior distribution over the latent variable and $p_\theta(\mathbf{x} | \mathbf{z})$ is the conditional likelihood that characterizes how the observed data is generated from the latent variable. In the VAE, the conditional likelihood is parameterized by a neural network called the decoder, whose parameters are denoted by $\theta$. The VAE also comes with an inference model $q_\phi(\mathbf{z} | \mathbf{x})$, which approximates the intractable posterior distribution of the latent variable. This inference model is parametrized by a second neural network, called the encoder, whose parameters are denoted by $\phi$. The inference model allows us to extract the latent variable $\mathbf{z}$ from an observed data vector $\mathbf{x}$, and the generative model allows us to generate $\mathbf{x}$ from $\mathbf{z}$. The parameters of both the inference (i.e., encoder) and generative (i.e., decoder) models are efficiently and jointly learned by maximizing a lower bound of the training data log-likelihood called the evidence lower-bound (ELBO), which is central in variational methods for inference and learning in probabilistic graphical models \citep{neal1998view,jordan1999introduction}.

\label{page:deux}
The VAE enables deep unsupervised representation learning in a Bayesian framework. Typically, a standard Gaussian distribution is chosen for the prior distribution over the latent variable: $p(\mathbf{z}) = \mathcal{N}(\mathbf{z}; \mathbf{0}, \mathbf{I})$. This choice encourages the independence of the different dimensions in the learned representation, and it is considered a key factor contributing to VAE's potential for disentanglement. However, it has been observed that vanilla VAEs exhibit limited disentanglement capability, especially when dealing with complex datasets. To address this challenge, substantial efforts have been made to enhance disentanglement by introducing implicit or explicit inductive biases in the model and/or in the learning algorithm.
Early methods for disentanglement using VAEs focused on modifying the evidence lower bound objective function \citep{higgins2016beta, chen2018isolating, kim2018disentangling}. Since unsupervised disentanglement in a generative model is impossible without incorporating inductive biases on both models and data \citep{locatello2019challenging}, new approaches are oriented towards weakly-supervised \citep{locatello2020weakly, sadok2023learning} or semi-supervised learning~\citep{klys2018learning}. Because of their flexibility in modeling complex data, VAEs have been extended to various types of data, including multimodal or sequential data.

VAEs have gained significant interest in modeling multimodal data due to their several advantages compared to other generative models, especially generative adversarial networks (GANs) \citep{goodfellow2014generative}. VAEs are equipped with encoder and decoder models, resulting in a more stable and faster training process than GANs, which makes them well-suited for multimodal generative modeling \citep{suzuki2022survey}.
Several approaches have been developed to learn a joint latent space for multiple heterogeneous input data.  For instance, PoE-VAE \citep{wu2018multimodal} adopts the product of experts (PoE) \citep{hinton2002training} to model the posterior distribution of multimodal data while MoE-VAE \citep{shi2019variational} uses a mixture of experts (MoE). Another approach \citep{sutter2021generalized} combines these methods for improved data reconstruction. Nevertheless, limitations have been demonstrated and formalized for these methods \citep{daunhawer2021limitations}. For example, multimodal VAE models often produce lower-quality reconstructions compared to unimodal VAE models, particularly for complex data. To address this issue and achieve better inference, many multimodal generative models now use hierarchical approaches to disentangle joint information from modality-specific information \citep{hsu2018disentangling, lee2020private, sutter2020multimodal}.

Another area where VAE models have seen significant progress is in the modeling of sequential data, where the latent and/or observed variables evolve over time. Dynamical VAEs (DVAEs) \citep{girin2021dynamical} aim to tackle high-dimensional complex data exhibiting temporal or spatial correlations using deep dynamical Bayesian networks. Recurrent neural networks are often used for this purpose, and a wide range of methods have been developed that differ in their inference and generative model structures. These DVAE models have two points in common when modeling sequential data: (i) unsupervised training is preserved, and (ii) the structure of the VAE is maintained; this means that the inference and generative models are jointly learned by maximizing a lower bound of the log-marginal likelihood (ELBO). Of particular interest to the present paper is the disentangled sequential autoencoder (DSAE) \citep{li2018disentangled}, which separates dynamical from static latent information.

While many extensions of the VAE have been proposed to handle either multimodal or sequential data, none have been able to process both types of data simultaneously. This paper presents a novel approach for modeling multimodal and sequential data in a single framework, specifically applied to audiovisual speech data. We propose the first unsupervised generative model of multimodal and sequential data, to learn a hierarchical latent space that separates static from dynamical information and modality-common from modality-specific information. The proposed model, called Multimodal Dynamical VAE (MDVAE), is trained on an expressive audiovisual speech database and evaluated on three tasks: the transformation of audiovisual speech data, audiovisual facial image denoising, and audiovisual speech emotion recognition.

\section{Multimodal Dynamical VAE}
\label{sec:MDVAE}

This section presents the design and architecture of MDVAE. Initially, we motivate the structure of the MDVAE latent space from the perspective of audiovisual speech generative modeling. Subsequently, we formalize the MDVAE generative and inference models. Finally, we introduce a two-stage training approach for unsupervised learning of the MDVAE model.

\begin{table}[t]
\centering
\caption{Summary of the notations.}
\label{tab:notation}
\resizebox{1.0\linewidth}{!}{ 
\begin{tabular}{ll}
    \hline 
    Variable notation & Definition  \\ \hline \\
    $T, t$ & Sequence length and time/frame index \\
    $\xa \in \mathbb{R}^{d_a \times T}$ & Observed audio data sequence\\ 
    $\xv \in \mathbb{R}^{d_v \times T}$ & Observed visual data sequence\\ 
    $\w \in \mathbb{R}^{w}$ & Latent static audiovisual vector\\ 
    $\zav \in \mathbb{R}^{l_{av} \times T}$ & Latent dynamical audiovisual vectors\\ 
    $\za \in \mathbb{R}^{l_a \times T}$ & Latent dynamical audio vectors\\ 
    $\zv \in \mathbb{R}^{l_v \times T}$ & Latent dynamical visual vectors\\ 
    $\mathbf{z} = \{ \za, \zv, \zav, \w \}$ & Set of all latent variables\\
    $\mathbf{x} = \{ \xa, \xv\}$ & Set of all observations\\ \\
    \hline
\end{tabular}
}
\end{table}

\subsection{Motivation and notations}
\label{subsectio:motivation}
Our goal is to model emotional audiovisual speech at the utterance level, where a single speaker speaks and expresses a single emotion. Let $\{\xa, \xv\}$ denote the observed audiovisual speech data, where $\xa \in \mathbb{R}^{d_a \times T}$ is a sequence of audio features of dimension $d_a$, $\xv \in \mathbb{R}^{d_v \times T}$ is a sequence of observed visual features of dimension $d_v$, and $T$ is the sequence length. For the audio speech, features are extracted from the power spectrogram of the signal, and for the visual speech, features are extracted from the pre-processed face images. The feature extraction process will be further discussed below.

To motivate the structure of the generative model in MDVAE, let us reason about the latent factors involved in generating an emotional audiovisual speech sequence. First, we have the speaker’s identity and global emotional state that correspond to static and audiovisual latent factors. Indeed, these do not evolve with time at the utterance level, and they are shared between the two modalities as defined from both vocal and visual attributes (e.g., the average pitch and timbre of the voice and the visual appearance). Second, we have dynamical latent factors that are shared between the two modalities, so audiovisual factors that vary with time. This typically corresponds to the phonemic information carried by the movements of the speech articulators that are visible in the visual modality, namely the jaw and lips. Finally, we have dynamical latent factors that are specific to each modality. Visual-only dynamical factors include, for instance, facial movements that are not related to the mouth region and the head pose. Audio-only dynamical factors include the pitch variations, induced by the vibration of the vocal folds, and the phonemic information carried by the tongue movements, which is another important speech articulator that is not visible in the visual modality. 

This analysis of the latent factors involved in the generative process of emotional audiovisual speech suggests structuring the latent space of the MDVAE model by introducing the following latent variables: $\w \in \mathbb{R}^w$ is a static latent variable assumed to encode audiovisual information that does not evolve with time; $\zav \in \mathbb{R}^{l_{av} \times T}$ is a dynamical (i.e., sequential) latent variable assumed to encode audiovisual information that evolves with time; $\za \in \mathbb{R}^{l_a \times T}$ is a dynamical latent variable assumed to encode audio-only information; $\zv \in \mathbb{R}^{l_v \times T}$ is a dynamical latent variable assumed to encode visual-only information. A time/frame index $t \in \{1,2,...,T\}$ is added in subscript of dynamical variables to denote one particular frame within a sequence (i.e., $\xat$, $\xvt$, $\zat$, $\zvt$, $\zavt$). The above notations are summarized in Table~\ref{tab:notation}. 
Note that for the specific modeling of audiovisual speech data, it is not particularly relevant to introduce static latent variables that are specific to each modality. Indeed, as above-described, the static information is here assumed to encode the speaker's identity and global emotional state, which are fundamentally multimodal factors. Nevertheless, for other types of data, we can readily envision using two distinct static latent variables for each modality. The methodology developed in the subsequent sections could be straightforwardly extended to this case. 

In summary, the MDVAE model is a generative model of audiovisual speech data $\{\xa, \xv\}$ that involves four different latent variables $\{\w, \zav, \za, \zv\}$. In the latent space of MDVAE, we can dissociate the latent factors that are static ($\w$) from those that are dynamic ($\zav, \zv, \za$), and we can dissociate the latent factors that are shared between the modalities ($\w, \zav$) from those that are specific to each modality ($\za, \zv$). Note that a study by Gao and Shinkareva \citep{gao2021modality} recently showed that the human brain distinguishes between modality-common and modality-specific information for affective processing in a multimodal context. In the MDVAE model, we also introduce temporal modeling on top of this dichotomy regarding modality-common vs modality-specific information. Our objective is to learn a multimodal and dynamical VAE than can disentangle the above-mentioned latent factors in an unsupervised manner for the analysis and transformation of emotional audiovisual speech data. In the next subsections, we detail the generative and inference models of MDVAE and its two-stage training.

\subsection{Generative model}
\label{section:generative-model}

\begin{figure}[t]
    \centering
    \includegraphics[width=0.5\textwidth]{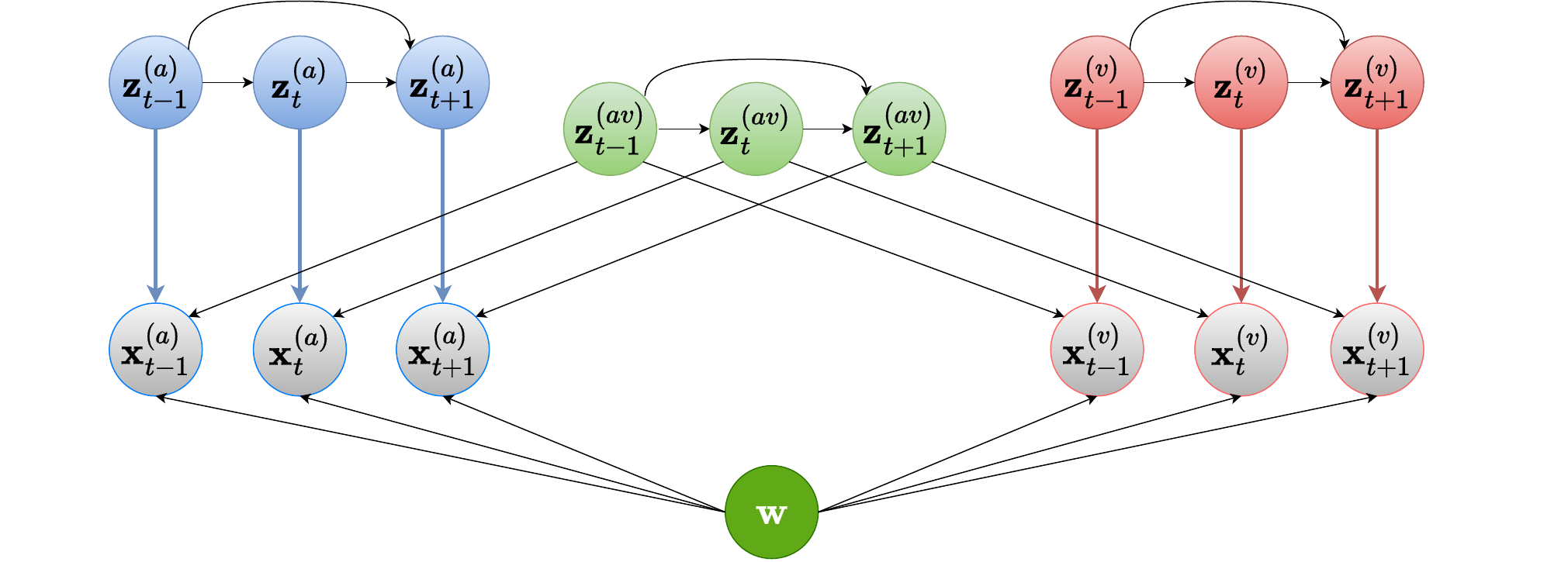}
    \caption{MDVAE generative probabilistic graphical model.}
    \label{fig:graph}
\end{figure}

The generative model of MDVAE is represented as a Bayesian network in Figure~\ref{fig:graph}, which also corresponds to the following factorization of the joint distribution of the observed and latent variables:
\begin{align}
p_\theta(\x, \z) &= p_\theta\left(\xa | \w, \zav, \za \right) p_\theta\left(\xv | \w, \zav, \zv \right) \nonumber \\
& \hspace{.75cm} \times p\left(\w\right) p_\theta\left(\zav\right) p_\theta\left(\za\right) p_\theta\left(\zv\right),
\label{eq:generative}
\end{align}
where $\mathbf{x} = \{ \xa, \xv\}$, $\mathbf{z} = \{ \za, \zv, \zav, \w \}$, and 

\begin{align}
p_\theta\left(\xa | \w, \zav, \za \right) &= \prod\limits_{t=1}^T p_\theta\left( \xat | \w, \zavt, \zat \right); \label{eq:generative_xa}
\end{align}

\begin{align}
p_\theta\left(\xv | \w, \zav, \zv\right) &= \prod\limits_{t=1}^T p_\theta\left( \xvt | \w, \zavt, \zvt \right); \label{eq:generative_xv}
\end{align}
\begin{align}
p(\zav) &= \prod\limits_{t=1}^T p_\theta\left(\zavt | \zavtt \right); \label{eq:generative_zav}
\end{align}
\begin{align}
p(\za) &= \prod\limits_{t=1}^T p_\theta\left(\zat | \zatt\right); \label{eq:generative_za}\\
p(\zv) &= \prod\limits_{t=1}^T p_\theta\left(\zvt | \zvtt \right). \label{eq:generative_zv}
\end{align}

\Eqref{eq:generative_xa} (resp. \eqref{eq:generative_xv}) indicates that, at time index $t$, the observed audio (resp. visual) speech vector $\xat$ (resp. $\xvt$) is generated from the audiovisual static latent variable ($\w$), the audiovisual dynamical latent variable at time index $t$ ($\zavt$), and the audio-only (resp. visual-only) dynamical latent variable at time index $t$ ($\zat$, resp. $\zvt$). In particular, we see that $\w$ is involved in the generation of the complete audiovisual speech sequence $(\xa, \xv)$. All latent variables are assumed independent, and the autoregressive structure of the priors for the dynamical variables in equations \plaineqref{eq:generative_zav}-\plaineqref{eq:generative_zv} is inspired by DSAE \citep{li2018disentangled}. Following standard DVAEs \citep{girin2021dynamical}, each conditional distribution that appears in a product over the time indices in equations~\plaineqref{eq:generative_xa}-\plaineqref{eq:generative_zv} is modeled as a Gaussian with a diagonal covariance, and its parameters are provided by deep neural networks (decoders) that take as input the variables after the conditioning bars. For the distributions in equations~\plaineqref{eq:generative_xa}-\plaineqref{eq:generative_xv}, the variance coefficients are fixed to one, while for the distributions in equations~\plaineqref{eq:generative_zav}-\plaineqref{eq:generative_zv}, the variance coefficients are learned. Standard feed-forward fully-connected neural networks can be used for parametrizing the conditional distributions over the observed audiovisual speech variables. The autoregressive structure of the priors over the latent dynamical variables requires the use of RNNs. Finally, the prior over the static latent variable $\w$ is a Gaussian with zero mean and identity covariance matrix. More details about the decoder network architectures can be found in \ref{appendix:architecture}.

\begin{figure}[t]
    \centering
    \includegraphics[width=0.5\textwidth]{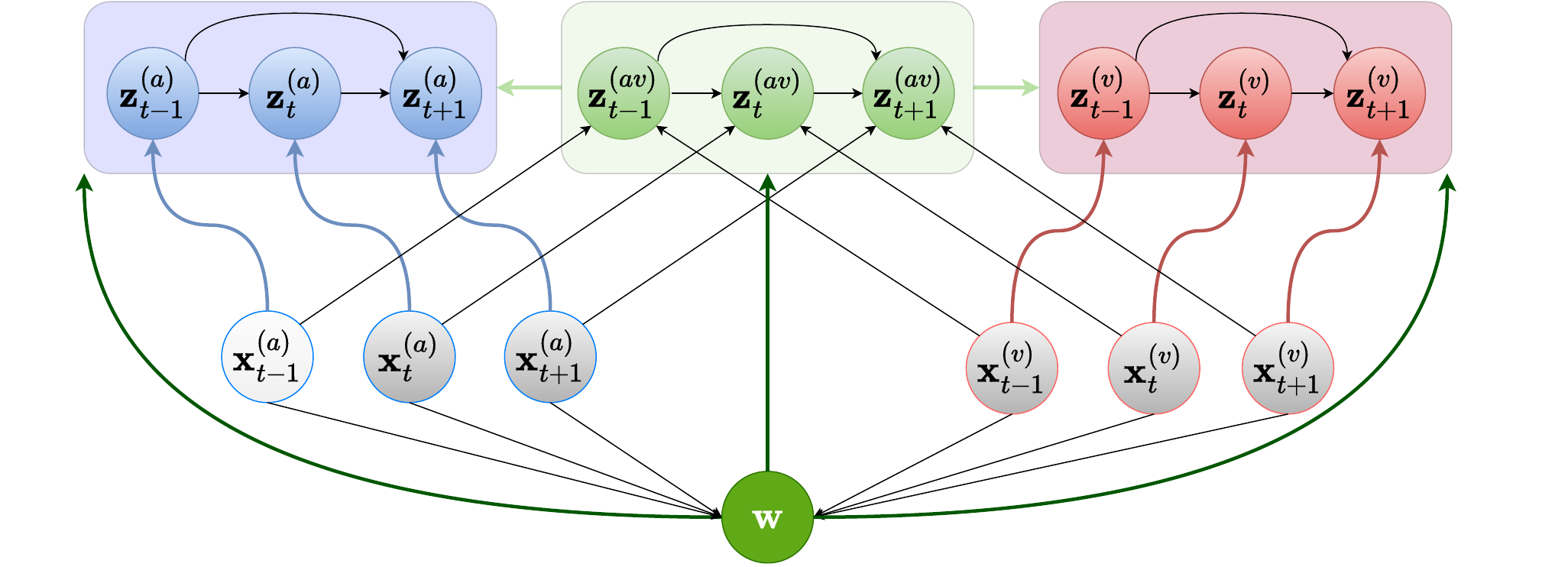}
    \caption{MDVAE inference probabilistic graphical model.}
    \label{fig:graph_inference}
\end{figure}

\subsection{Inference model}
\label{section:inference-model}

As in the standard VAE, the exact posterior distribution of the latent variables in the MDVAE model is intractable, we thus need to define an inference model $q_\phi\left(\z | \x\right) \approx p_\theta\left(\z | \x\right)$. However, it is not because the exact posterior distribution is intractable that we cannot look at the structure of the exact posterior dependencies. Actually, using the Bayesian network of the model, the chain rule of probabilities, and D-separation \citep{geiger1990identifying, bishop2006pattern}, it is possible to analyze how the observed and latent variables depend on each other in the exact posterior, and define an inference model with the same dependencies. An extensive discussion of D-separation in the context of DVAEs can be found in \citep{girin2021dynamical}. The Bayesian network corresponding to our MDVAE model is represented in Figure~\ref{fig:graph}. For this model, it is relevant to factorize the inference model as follows:
\begin{align}
\label{eq:inference}
& q_\phi\left(\z | \x\right) = q_\phi\left(\w | \xa, \xv\right)q_\phi\left(\zav | \xa, \xv, \w \right) \nonumber\\
\hspace{.25cm} & \times q_\phi\left(\za | \xa, \zav, \w \right)q_\phi\left(\zv | \xv, \zav, \w \right),
\end{align}
where
\begin{align}
\label{eq:inference-z_av}
q_\phi\left(\zav | \xa, \xv, \w \right) &= \prod\limits_{t=1}^T q_\phi\left(\zavt \mid \zavtt, \xatT, \xvtT, \w \right); \\
\label{eq:inference-z_a}
q_\phi\left(\za | \xa, \zav, \w \right) &= \prod\limits_{t=1}^T q_\phi\left(\zat \mid \zatt, \xatT, \zavt, \w \right); \\
\label{eq:inference-z_v}
q_\phi\left(\zv | \xv, \zav, \w \right) &= \prod\limits_{t=1}^T q_\phi\left(\zvt \mid \zvtt, \xvtT, \zavt, \w \right).
\end{align}
This factorization is consistent with the exact posterior dependencies between the latent and observed variables, i.e., no approximation was made as we followed the principle of D-separation. However, to lighten the inference model architecture, we choose to omit the non-causal dependencies on the observations in \eqref{eq:inference-z_av}, \eqref{eq:inference-z_a} and \eqref{eq:inference-z_v}. In these equations, we thus replace $\xatT$ by $\xat$ and $\xvtT$ by $\xvt$, and the equalities become approximations. In this inference model, $q_\phi\left(\w | \xa, \xv\right)$ and each conditional distribution that appears in a product over the time indices in equations~\plaineqref{eq:inference-z_av}-\plaineqref{eq:inference-z_v} is modeled as a Gaussian with a diagonal covariance, and its parameters (mean vector and variance coefficients) are provided by deep neural networks (encoders) that take as input the variables after the conditioning bars. In practice, the MDVAE encoder can be decomposed into four sub-encoders, each dedicated to the inference of a specific latent variable. Distinct conditioning variables are concatenated at the input of these sub-encoders depending on the structure of the corresponding inference model. For instance, when inferring $\w$ we concatenate $\xa$ and $\xv$ along the feature dimension.
More details about the encoder network architectures can be found in \ref{appendix:architecture}. 

\begin{figure*}[t]
    \centering
    \includegraphics[width=\textwidth]{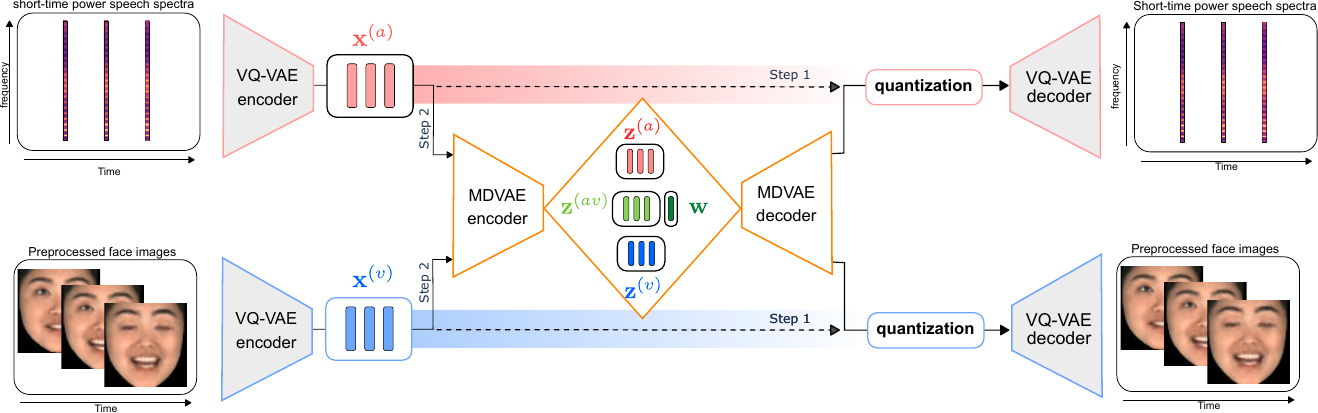}
    \caption{The overall architecture of VQ-MDVAE. During the first step of the training process, we learn a VQ-VAE independently on each modality, without any temporal modeling. During the second step of the training process, we learn the MDVAE model on the latent representation provided by the frozen VQ-VAE encoders, before quantization.}
    \label{fig:twostageMDVAE}
\end{figure*}

The probabilistic graphical model of MDVAE during inference is represented in Figure~\ref{fig:graph_inference}, corresponding to the factorization in \eqref{eq:inference}. It can be interpreted as follows: First, we infer the static audiovisual latent variable $\w$ from the observed audiovisual speech sequence, which corresponds to the computation of $q_\phi(\w | \xa, \xv)$. Next, we infer the audiovisual dynamical latent variable $\zav$ from the previously inferred variable $\w$ and the observed audiovisual speech, which corresponds to the computation of $q_\phi(\zav | \xa, \xv, \w )$. Indeed, we need the static audiovisual information to infer the dynamical audiovisual information from the audiovisual speech observations. Finally, we infer the audio-only (resp. visual-only) dynamical latent variables $\za$ (resp. $\zv$) from the audio (resp. visual) speech observations $\xa$ (resp. $\xv$) and the previously inferred audiovisual latent variables $\w$ and $\zav$, which corresponds to the computation of $q_\phi(\za | \xa, \zav, \w )$ (resp. $q_\phi(\zv | \xv, \zav, \w )$). This is logical, as to infer the latent information that is specific to one modality, we require the observations of that modality and also the latent information that is shared with the other modality, which is captured by $\w$ and $\zav$.

\subsection{Training}
\label{subsec:training}
As in standard (D)VAEs \citep{kingma2014auto,rezende2014stochastic, girin2021dynamical}, learning the MDVAE generative and inference model parameters consists in maximizing the evidence lower-bound (ELBO):
\begin{align}
\label{eq:elbo}
\mathcal{L}(\theta, \phi) = \mathbb{E}_{q_\phi(\z \mid \x)}\left[\ln p_\theta(\x \mid \z)\right] - D_{\text{KL}}\left(q_\phi(\z \mid \x) \parallel p_\theta(\z)\right).
\end{align}
where $D_{\text{KL}}$ is the Kullback-Leibler divergence, defined by $D_{\text{KL}}(q \parallel p) =  \mathbb{E}_q[\ln q - \ln p]$. The first term in \eqref{eq:elbo} is the reconstruction accuracy term, which aims to maximize la data log-likelihood over a training dataset. These input and output data can take any form, including raw images for the visual modality and speech power spectra for the audio modality, or can be replaced by any representation from another pre-trained model. The second term is the latent space regularization term, which encourages the latent variables to conform to the prior distribution. 
Using equations~\plaineqref{eq:generative} and \plaineqref{eq:inference}, the ELBO can be further developed as follows:

\begin{align}
    \label{eq-appendix:elbo}
     & \mathcal{L}(\theta, \phi)=\mathbb{E}_{q_\phi(\z|\x)}\left[\ln p_\theta\left(\xa, \xv \mid \w, \zav, \za, \zv\right) \right]\nonumber\\
     &\hspace{.2cm} - D_{KL}(q_\phi(\w|\xa, \xv) \parallel p_\theta(\w)) \nonumber\\
     &\hspace{.2cm} -\mathbb{E}_{q_\phi(\z|\x)}\left[D_{KL}\left(q_\phi\left(\zav|\xa, \xv, \w\right) \parallel p_\theta\left(\zav\right)\right)  \right]\nonumber\\
     &\hspace{.2cm} -\mathbb{E}_{q_\phi(\z|\x)}\left[D_{KL}\left(q_\phi\left(\za|\xa, \w, \zav\right) \parallel p_\theta\left(\za\right)\right)  \right]\nonumber\\
     &\hspace{.2cm} -\mathbb{E}_{q_\phi(\z|\x)}\left[D_{KL}\left(q_\phi\left(\zv|\xv, \w, \zav\right) \parallel p_\theta\left(\zv\right)\right)  \right].
\end{align}

\subsection{Two-stage training}
\label{sec:two_stage}

Unlike GANs \citep{goodfellow2014generative}, VAEs often produce poor reconstructions that lack realism, and this also affects the generation of new data. Improving the quality of VAE reconstruction or generation is an active area of research. One issue with VAE is that using an information bottleneck in combination with a pixel-wise reconstruction error can result in blurry, unrealistic images. This problem also exists with the audio modality, where VAE-generated sound is often unnatural, mainly when using a time-frequency representation. 
To address this problem, several solutions have been proposed. One approach is to combine VAEs and GANs, where the discriminator replaces the standard reconstruction error and provides improved realism \citep{larsen2016autoencoding}. Another solution is to build a hierarchical VAE, with a more complex structure for the latent space \citep{vahdat2020nvae}. Other methods incorporate regularization techniques, such as using a perceptual loss for the image modality, to ensure that VAE outputs have similar deep features to their corresponding inputs \citep{pihlgren2020improving,hou2019improving}.
In this work, we focus on using a vector quantized VAE (VQ-VAE) model \citep{van2017neural}, which is a deterministic autoencoder with a discrete latent space. In the VQ-VAE, the continuous latent vector provided by the encoder is quantized using a discrete codebook before being fed to the decoder network. The codebook is jointly learned with the network architecture. The VQ-VAE model has been shown to produce higher-quality generations than VAEs or GANs~\citep{razavi2019generating}.
Therefore, as illustrated in Figure~\ref{fig:twostageMDVAE}, we propose a two-stage training approach of the MDVAE model to improve its reconstruction and generation quality. 

The first stage involves learning a VQ-VAE model independently on the visual and audio modalities and without temporal modeling. 
The training procedure of the VQ-VAEs, including the loss functions, is the same as originally proposed in \citep{van2017neural}, using an exponential moving average for the codebook updates. The VQ-VAE loss function includes a reconstruction term, which corresponds to the pixel-wise mean squared error for the visual modality and to the Itakura-Saito divergence \citep{fevotte2009nonnegative} for the audio modality.
In the second stage, we learn the MDVAE model on the continuous representations obtained from the pre-trained VQ-VAE encoders before quantization, instead of working directly on the raw audiovisual speech data. The disentanglement between static versus dynamic and modality-specific versus audiovisual latent factors occurs during this second training stage. This is because the VQ-VAEs are learned independently on each modality and without temporal modeling. To reconstruct the data, the continuous representations from the MDVAE are quantized and decoded by the pre-trained VQ-VAE decoders. This approach will be referred to as VQ-MDVAE in the following.

The first stage of this two-stage approach can be seen as learning audiovisual speech features in an unsupervised manner using a VQ-VAE. This feature extraction procedure is pseudo-invertible, as we can go from the raw data to the features with the VQ-VAE encoder and from the features to the raw data with the VQ-VAE decoder. 

Beyond the reconstruction quality of the audiovisual speech data, another interest of the proposed two-stage training procedure is that it allows us to distribute the GPU memory usage between the two training stages, which is particularly useful when working with limited computational resources. During the first training stage, we train \emph{small models} (fully convolutional VQ-VAEs) on \emph{high-dimensional data}, corresponding to the raw audio and visual speech data. Processing modalities and time frames independently along with defining models of reasonable size allows us to efficiently manage the GPU memory to create large batches containing the raw audio or visual speech data. During the second training stage, we train a \emph{large model} (MDVAE) on \emph{low-dimensional data}, corresponding to the compressed audio and visual latent representations provided by the pre-trained VQ-VAEs before quantization. The effective compression of the audiovisual speech data using the VQ-VAEs allows us to increase the model capacity for the second training stage, which is required by the complexity of the learning task, and at the same time to keep a sufficiently large batch size. In summary, the two-stage training strategy decomposes the difficult problem of learning from high-dimensional multimodal and sequential data into two smaller sub-problems. The first sub-problem deals with compressing the data and ensuring a good reconstruction quality, and the second sub-problem deals with modality fusion and temporal modeling. Overall, this two-stage training leads to good reconstruction quality, efficient memory management, and accelerated training speed. Note that from a practical perspective, one could learn the VQ-VAE and MDVAE models jointly, from scratch, given sufficient computational resources.

\section{Experiments on audiovisual speech}
\label{sec:Experiments}
This section presents three sets of experiments conducted with the VQ-MDVAE model for audiovisual speech processing. First, we analyze qualitatively and quantitatively the learned representations by manipulating audiovisual speech sequences in the MDVAE latent space. Second, we explore the use of the VQ-MDVAE model for audiovisual facial image denoising, showing that the model effectively exploits the audio modality to reconstruct facial images where the mouth region is corrupted. Finally, we show that using the static audiovisual latent representation learned by the VQ-MDVAE model leads to state-of-the-art results for audiovisual speech emotion recognition.

\subsection{Expressive audiovisual speech dataset}
\label{sec:audiovisual-dataset}

The VQ-MDVAE model is trained on the multi-view emotional audiovisual dataset (MEAD) \citep{wang2020mead}. It contains talking faces comprising 60 actors and actresses speaking with eight different emotions at three levels of intensity. We keep only the frontal view for the visual modality. 75\%, 15\%, and 10\% of the dataset are used respectively for the training, validation, and test, with different speakers in each split. This corresponds to approximately 25h, 5h, and 3h of audiovisual speech, respectively. 
For the visual modality, face images in the MEAD dataset are cropped, resized to a 64x64 resolution, and aligned using Openface \citep{baltruvsaitis2016openface}. For the audio modality, power spectrograms are computed using the short-time Fourier transform (STFT). The STFT parameters are chosen such that the audio frame rate is equal to the visual frame rate (30 fps), which leads to an STFT analysis window length of 64 ms (1024 samples at 16 kHz) and a hop size of $52.08\%$ of the window length. 

\subsection{Training VQ-MDVAE}

The architecture of the MDVAE and VQ-VAE models are described in detail in \ref{appendix:architecture}. This section only provides an overview of the training pipeline.

The pre-processed facial images and the power spectrograms are used to train the visual and audio VQ-VAEs, respectively. The two VQ-VAEs do not include any temporal model, i.e., the audio and visual frames of an audiovisual speech sequence are processed independently. The VQ-VAE for the visual modality takes as input and outputs an RGB image of dimension $64 \times 64 \times 3$. This image is mapped by the encoder to a latent representation corresponding to a 2D grid of $8 \times 8$ codebook vectors of dimension $32$. The visual codebook contains a total number of $512$ vectors. The VQ-VAE for the audio modality takes as input and outputs a speech power spectrum of dimension $513$. This power spectrum is mapped by the encoder to a latent representation corresponding to a 1D grid of $64$ codebook vectors of dimension $8$. The audio codebook contains a total number of $128$ vectors.
The VQ-VAEs consist of convolutional layers for both the visual and audio modalities. Since the quantization operation is non-differentiable, the codebooks for each modality are learned using the stop gradient trick \citep{van2017neural}. 

The audio and visual observed data $\xa \in \mathbb{R}^{d_a \times T}$ and $\xv \in \mathbb{R}^{d_v \times T}$ that are used to train the MDVAE model are taken from the flattened output of the pre-trained and frozen VQ-VAE encoders before quantization, with $d_a = 512$ ($64 \times 8$) and $d_v = 2048$ ($8 \times 8 \times 32$). The sequence length is fixed to $T=30$ for training. The MDVAE model is composed of dense and recurrent layers. The dimensions of the latent variables in the VQ-MDVAE model are as follows: the static latent vector ($\w \in \mathbb{R}^w$) has a dimension of $w = 84$, the audiovisual dynamical latent vectors ($\zav \in \mathbb{R}^{l_{av} \times T}$) have a dimension of $l_{av} = 16$, and both the audio and visual dynamical latent vectors ($\zv \in \mathbb{R}^{l_{v} \times T}, \za \in \mathbb{R}^{l_{a} \times T}$) have a dimension of $l_{v} = l_{a} = 8$. The models are trained using the Adam optimizer \citep{DBLP:journals/corr/KingmaB14}. 

\subsection{Analysis-resynthesis}
\label{sec:analysis-resynthesis}
We first present the results of an \emph{analysis-resynthesis} process on the audiovisual speech data. The \emph{analysis} step involves performing inference on audiovisual speech sequences that were not seen during training to obtain the latent vectors, while the \emph{resynthesis} step involves generating the sequence from the obtained latent vectors without any modification, with the goal of faithfully reconstructing the input sequence.

\noindent\textbf{Methods}\hspace{0.20cm} For this experiment, we compare MDVAE and VQ-MDVAE to VQ-VAE \citep{van2017neural} and DSAE \citep{li2018disentangled}, which are unimodal generative models. DSAE also includes a temporal model that separates sequential information from static information. The VQ-VAE does not include any temporal model. The VQ-VAE and DSAE are both trained separately on the audio and visual modalities. For a fair comparison, we consider the original DSAE and its improved version VQ-DSAE obtained by training the model in two stages like VQ-MDVAE (see Section~\ref{sec:two_stage}). This experimental comparison therefore corresponds to an ablation study: If we take VQ-MDVAE and remove the multimodal modeling we obtain VQ-DSAE. If we further remove the temporal model we obtain VQ-VAE. It will also allow us to assess the impact of the proposed two-stage training process on both DSAE and MDVAE.

\noindent\textbf{Evaluation metrics}\hspace{0.20cm} The average quality performance for the speech and visual modalities is evaluated using the MEAD test dataset.
Four metrics are used to assess the quality of the resynthesized audio speech data: 
\begin{itemize}[leftmargin=1em]
    \item The Short-Time Objective Intelligibility (STOI) measure is an intrusive metric (i.e., it requires the original reference speech signal) that assesses how intelligible the resynthesized speech is \citep{taal2010short};
    \item The Perceptual Evaluation of Speech Quality (PESQ) measure is an intrusive metric that evaluates the perceived quality of the resynthesized speech \citep{rix2001perceptual}. It accounts for factors like distortion, noise, and other artifacts that can affect the overall perceived quality;
    \item The Scale-Invariant Signal-to-Distortion Ratio (SI-SDR) is an intrusive metric defined as the power ratio between the original speech signal and the distortion caused by the resynthesis process \citep{le2019sdr}; it is made invariant to signal amplitude rescaling;
    \item MOSnet is a learning-based non-intrusive metric that predicts human-rated quality scores for speech \citep{lo2019mosnet}.
\end{itemize}
Four metrics are also used to assess the quality of the resynthesized visual data: 
\begin{itemize}[leftmargin=1em]
    \item The Mean Square Error (MSE) computes the average squared difference between the pixel values of the original and resynthesized visual data;
    \item The Peak Signal-to-Noise Ratio (PSNR) considers both the image fidelity and the level of noise or distortion introduced during resynthesis;
    \item The Spatial Correlation Coefficient (SCC) evaluates how well the structures and patterns in the images match using the correlation;
    \item The Structural Similarity Index Measure (SSIM) assesses the structural similarity between the original and resynthesized images. It takes into account luminance, contrast, and structure, providing a comprehensive measure of image quality \citep{wang2004image}.
\end{itemize}

\noindent\textbf{Discussions}\hspace{0.20cm}
Tables~\ref{tab:audio-quality} and~\ref{tab:visual-quality} respectively show the reconstruction quality of the audio and visual modalities for this \emph{analysis-resynthesis} experiment. The proposed VQ-MDVAE method outperforms MDVAE alone, as evidenced by the improvement of 0.03, 0.47, 1.19, and 4.64 for STOI, PESQ, MOSnet, and SI-SDR, respectively, for the audio modality. Similarly, for the visual modality, VQ-MDVAE yields a gain of 6.5, 0.1, and 0.26 for PSNR, SCC, and SSIM, respectively. These results validate the proposed two-step training approach, demonstrating a significant improvement in reconstruction quality. This is confirmed when comparing the results of DSAE with those of VQ-DSAE.
In addition, for both modalities, MDVAE and VQ-MDVAE outperform DSAE and VQ-DSAE, respectively.
However, the proposed method (VQ-MDVAE) shows a decrease in reconstruction quality compared to using the VQ-VAE alone, especially for the PESQ metric. This can be attributed to the fact that the VQ-MDVAE, with its temporal dependencies, acts as a temporal filter. Despite this, we can leverage these temporal dependencies and the hierarchy provided by the MDVAE model for other applications, as discussed in the following sections.

\begin{figure}[t!]
\begin{minipage}{0.5\textwidth}
\begin{minipage}{1\textwidth}
\captionof{table}{Speech performance of the MDVAE model tested in the \emph{analysis-resynthesis} experiment. The STOI, PESQ, and MOSnet scores are averaged over the test subset of the MEAD dataset.}
\label{tab:audio-quality}
\resizebox{1.0\linewidth}{!}{ 
    \begin{tabular}{ccccc}
        \toprule
        Method         & \multicolumn{1}{c}{STOI $\uparrow$}  & \multicolumn{1}{c}{PESQ $\uparrow$}  & MOSnet  $\uparrow$ & SI-SDR $\uparrow$  \\ \hline
        VQ-VAE-audio          & 0.91 $\tiny \pm 0.02$               & 3.49 $\tiny \pm 0.25$               & 3.60 $\tiny \pm 0.15$ & 6.67 $\tiny \pm 1.18$\\ 
        DSAE-audio     & 0.79 $\tiny \pm 0.05$               & 2.10 $\tiny \pm 0.31$               & 1.88 $\tiny \pm 0.30$ & -1.20 $\tiny \pm 1.58$\\ 
        MDVAE     & 0.82 $\tiny \pm 0.03$               & 2.43 $\tiny \pm 0.28 $               & 2.35 $\tiny \pm 0.18$ & 2.21 $\tiny \pm 1.30$\\ \hline
        VQ-DSAE-audio     & 0.84 $\tiny \pm 0.03$               & 2.12 $\tiny \pm 0.24$               & 3.05 $\tiny \pm 0.20$ & 6.12 $\tiny \pm 1.10$\\ 
        VQ-MDVAE & 0.85 $\tiny \pm 0.04$               & 2.90 $\tiny \pm 0.23$ & 3.54 $\tiny \pm 0.20$  & 6.85 $\tiny \pm 1.15$\\ \bottomrule
    \end{tabular}
}
\end{minipage}
\begin{minipage}{1\textwidth}
\vspace{0.5cm}
\captionof{table}{Visual performance of the MDVAE model tested in the \emph{analysis-resynthesis} experiment. The MSE, PSNR, SCC and SSIM scores are averaged over the test subset of the MEAD dataset.}
\label{tab:visual-quality}
\resizebox{1.0\linewidth}{!}{ 
    \begin{tabular}{ccccc}
        \toprule
        Method         & \multicolumn{1}{c}{MSE $\downarrow$}  & \multicolumn{1}{c}{PSNR $\uparrow$}  & SCC  $\uparrow$    & SSIM  $\uparrow$    \\ \hline
        VQ-VAE-visual          & 0.0016 $\tiny \pm 0.0002$               & 27.2 $\tiny \pm 0.70$               & 0.70 $\tiny \pm 0.01$ & 0.85 $\tiny \pm 0.01$ \\ 
        DSAE-visual     & 0.023 $\tiny \pm 0.03$               & 15.8 $\tiny \pm 2.9$               & 0.58 $\tiny \pm 0.07$ & 0.47 $\tiny \pm 0.03$ \\
        MDVAE  & 0.010 $\tiny \pm 0.008$ &  20.3 $\tiny \pm 1.3$ & 0.62 $\tiny \pm 0.03$  & 0.58 $\tiny \pm 0.03$ \\ \hline
        VQ-DSAE-visual     & 0.0018 $\tiny \pm 0.0005$               & 25.3 $\tiny \pm 1.23$               & 0.70 $\tiny \pm 0.01$ & 0.82 $\tiny \pm 0.04$ \\
        VQ-MDVAE & 0.0017 $\tiny \pm 0.0007$ & 26.8 $\tiny \pm 0.72$ & 0.72 $\tiny \pm 0.01$ & 0.84 $\tiny \pm 0.02$ \\ \bottomrule
    \end{tabular}
}
\end{minipage}
\end{minipage}
\end{figure}

\begin{figure*}[t]
    \centering
    \includegraphics[width=\textwidth]{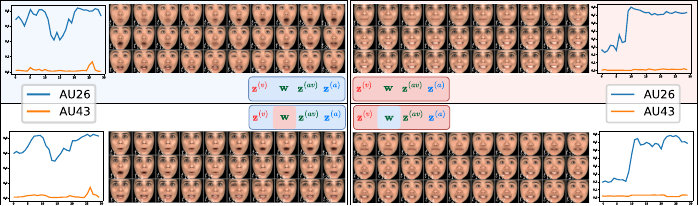}
    \caption{Visual sequences generated using the \emph{analysis-transformation-synthesis} experiment. The top two sequences depict original image sequences of two distinct individuals, while the bottom two sequences were generated by swapping the latent variable $\w$ between the two original sequences.}
    \label{fig:qualitative-visual-1}
\end{figure*}

\begin{figure*}[ht]
    \centering
    \includegraphics[width=\textwidth]{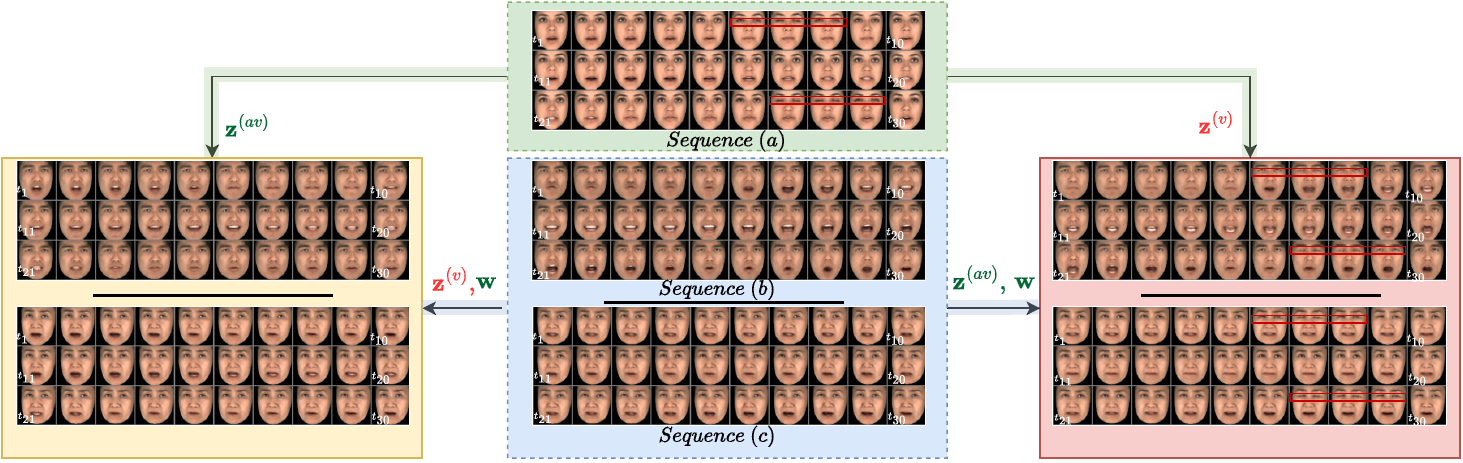}
    \caption{This figure demonstrates the qualitative significance of each latent space for visual data using the \emph{analysis-transformation-synthesis} experiment. The sequences in the yellow box (left) were generated using $\zav$ from sequence (a) and ${\zv, \w}$ from sequences (b) and (c). The sequences in the red box (right) were generated using $\zv$ from the sequence (a), and ${\zav, \w}$ from sequences (b) and (c).}
    \label{fig:qual-visual-1}
\end{figure*}

\begin{figure}[t!]
    \centering
    \includegraphics[width=0.48\textwidth]{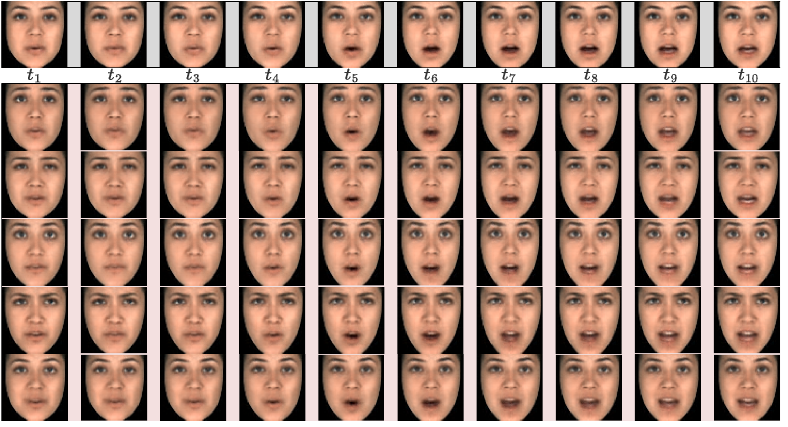}
    \caption{The first row represents a sequence of face images for an individual whose emotion is neutral. The rows below are generated with VQ-MDVAE, keeping all the dynamical latent variables of the first sequence and replacing the static latent variable with that of sequences from the same person but with different emotions (from top to bottom: fear, sad, surprised, angry, and happy).}
    \label{fig:qual-emotion-0}
\end{figure}

\subsection{Analysis-transformation-synthesis}
\label{sec:Analysis-transformation-synthesis}
This section aims to analyze the latent representations learned by the MDVAE model. We want to study what high-level characteristics of audiovisual speech are encoded in the different latent variables of the model.
The experiments involve exchanging latent variables between a sequence named (A) and sequences named (B) through an \emph{analysis-transformation-synthesis process}. The analysis step involves performing inference separately on two audiovisual speech sequences (A) and (B). Then, the values of certain latent variables from (A) are replaced with the values of the same latent variable from (B). Finally, the output sequence is reconstructed from the combined set of latent variables. The resulting sequence is expected to be a mixed sequence whose features correspond to sequence (A) for the unmodified latent variables and sequence (B) for the modified latent variables.

\begin{figure*}[h!]
    \centering
    \includegraphics[width=\textwidth]{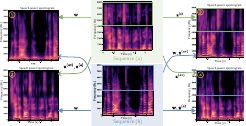}
    \caption{Audio spectrograms generated from \emph{analysis-transformation-synthesis} between sequence (a) in green and sequence (b) in blue. The spectrograms (1), (2), (3), and (4) are synthesized by swapping latent variables between sequence (a) and sequence (b). The black dotted line corresponds to the pitch contour.}
    \label{fig:qual-audio}
\end{figure*}

\subsubsection{Qualitative results}
\label{qualitative-parts}

\noindent\textbf{Visual modality}\hspace{0.20cm}
Figure~\ref{fig:qualitative-visual-1} illustrates visual sequences generated using the \emph{analysis-transformation-synthesis} method, each accompanied by two curves representing the intensity of two facial action units (AUs), namely jaw drop ($AU_{26}$) and eyes closed ($AU_{43}$), plotted as a function of the frame index. AUs are the smallest components of facial expression, involving coordinated contractions of facial muscles that produce recognizable and measurable changes in the face \citep{ekman1978facial}. These AUs were extracted from the visual sequences using Py-Feat \citep{muhammod2019pyfeat}.
The top two sequences depict original visual sequences of different subjects exhibiting varying facial expressions. Conversely, the bottom sequences display the results when the variable $\w$ values are swapped between the two original sequences.
We observe that the bottom-left sequence has the same facial movements as the top-left sequence, but the speaker identity is that of the top-right sequence. The curves of $AU_{43}$ and AU26 for the bottom-left sequence are similar to those of the top-left sequence. A noticeable blink of the eyes occurs between frames 26 and 28, which is depicted by a peak in the $AU_{43}$ curve. Similarly, the bottom-right sequence has the same facial movements as the top-right sequence, but the speaker identity is that of the top-left sequence. This disentanglement of dynamic facial movements from static speaker identity reveals that $\w$ encodes the visual identity of the speaker, among other information.

Figure~\ref{fig:qual-visual-1} illustrates what other latent variables encode using the \emph{analysis-transformation-synthesis} method. The figure shows three sequences of visual data, labeled as sequence \textit{(a)} in the green box and sequences \textit{(b)} and \textit{(c)} in the blue box.
First, two sequences on the left are reconstructed by combining $\zav$ of sequence (a) with $\w$ and $\zv$ of sequences (b) and (c). The speaker identity of sequences (b) and (c) is preserved in the output sequences, but the movement of the lips follows that of sequence (a). This shows that $\zav$ encodes the lip movement.
Second, two other sequences on the right are reconstructed by combining $\zv$ of sequence (a) with $\w$ and $\zav$ of sequences (b) and (c). The speaker identity and the movement of the lips of sequences (b) and (c) are preserved in the output sequences, but the movement of the eyes and eyelids (e.g., the blink of the eyes, as seen in the red rectangle) follows that of sequence (a). This indicates that $\zv$ encodes eye and eyelid movements. It also appears that the head orientation in the bottom right output sequence is different from that of the original sequence (c), which was not the case for the bottom left output sequence. This indicates that $\zv$ also encodes the head pose. From this example, we can also confirm that $\w$ encodes the speaker's identity.

Figure~\ref{fig:qual-emotion-0} shows that $\w$ also encodes the global emotional state. Each line in the figure is a reconstruction created by combining the dynamical latent variables of the sequence labeled as neutral in terms of emotion (first row) with $\w$ of other sequences of the same person labeled with different emotions (from top to bottom: fear, sad, surprised, angry, and happy). The emotion changes between the different rows, but the visual dynamics remain the same as in the first row, indicating that the static audiovisual variable $\w$ encodes both the identity and the global emotion in the input sequence. 

\noindent\textbf{Audio modality}\hspace{0.20cm} As for the visual modality, Figure~\ref{fig:qual-audio} illustrates audio sequences (speech power spectrograms) generated using the \emph{analysis-transformation-synthesis} method. In this figure, sequence (a) (green box) represents the power spectrogram and the pitch contour of a speech signal spoken by a male speaker, and sequence (b) (blue box) represents the power spectrogram and the pitch contour of a speech signal spoken by a female speaker. The pitch contour is extracted using CREPE \citep{kim2018crepe}. The generated spectrogram (1) (top left) is derived from $\w$ of sequence (a), and the dynamical latent variables $\zav, \za$ of sequence (b). Comparing the resulting spectrogram with that of sequence (b), we can deduce that they have the same phonemic structure, but the pitch has been shifted downwards, as can be seen from the pitch contour and the spacing between the harmonics. Similarly, the reconstructed spectrogram (2) (bottom left) is derived from $\w$ of sequence (b), and the dynamical latent variables $\zav, \za$ are from the sequence (a). Here, we notice that the pitch shifts upwards while preserving the phonemic structure of sequence (a). Therefore, the static latent variable $\w$ encodes the average pitch value related to the speaker's identity. The generated spectrograms (3) (top right) and (4) (bottom right) reveal that the dynamical latent variables $\za$ and $\zav$ have distinct roles in capturing the phonemic content. Specifically, $\za$ predominantly captures the high frequency, while $\zav$ encodes the low frequency, which also corresponds to the lower formants. This finding is noteworthy as research has shown that the lower formants are highly correlated with the lip configuration \citep{arnela2016influence}. Moreover, it is particularly interesting that the two correlated factors (lower formants and lip movements) are found in the same latent dynamical variable, $\zav$, especially since the MDVAE was trained in an unsupervised manner.

\noindent\textbf{Additional qualitative results}\hspace{0.20cm} Additional qualitative results, such as audiovisual animations, analysis-transformation-synthesis, interpolation on the static latent space, and audiovisual speech generation conditioned on specific latent variables, can be found at \url{https://samsad35.github.io/site-mdvae/}.

\subsubsection{Quantitative Results}
The aim of this section is to complement the above qualitative analysis with quantitative metrics by measuring the ability of the VQ-MDVAE model to modify facial and vocal attributes through manipulations of the different latent variables.

\begin{figure}[t]
    \centering
    \includegraphics[width=0.48\textwidth]{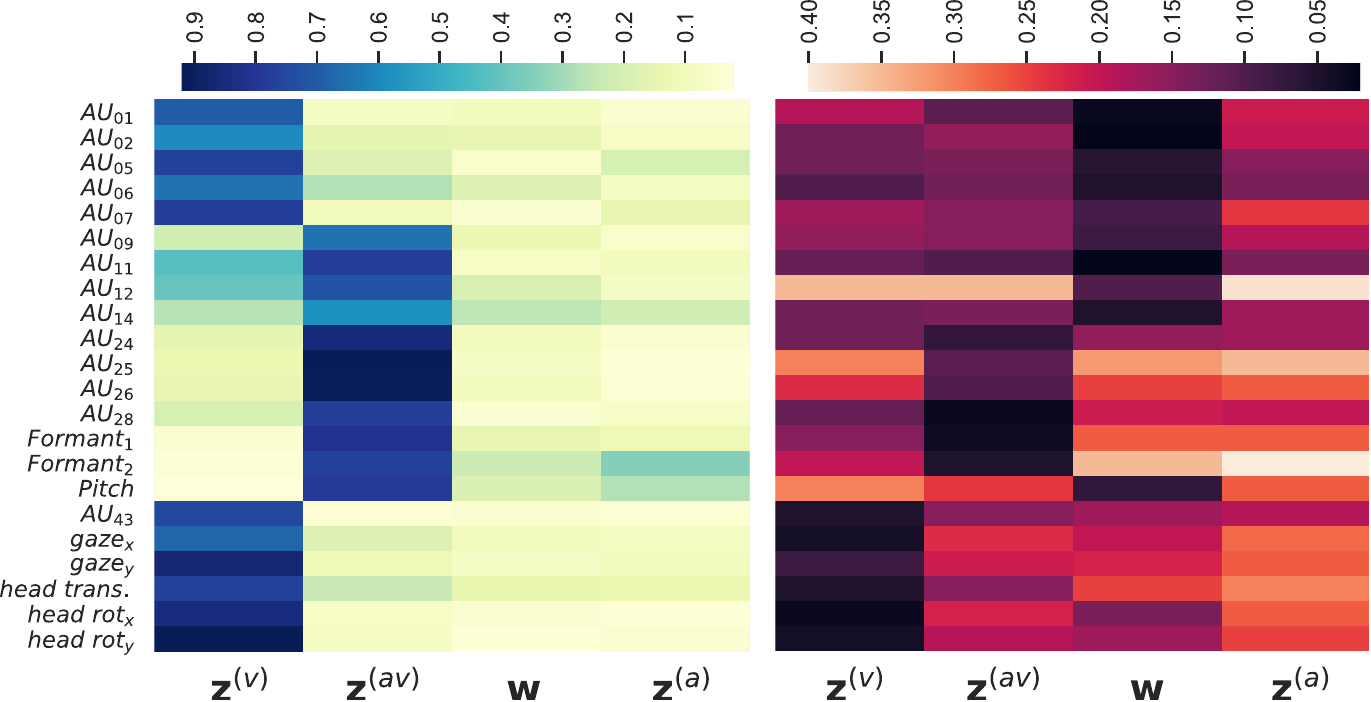}
    \caption{Relationship between the audio/visual attributes and the latent variables of VQ-MDVAE. (left) Pearson correlation coefficient (PCC), (right) mean absolute error (MAE).}
    \label{fig:corr-mae}
\end{figure}

\noindent\textbf{Experimental setup and metrics}\hspace{0.20cm}
The evaluation protocol for this experiment involves using a sequence (labeled as (A)) and 50 other sequences selected randomly from the test dataset (labeled as (B)). 
The protocol is based on the \emph{analysis-transformation-synthesis} framework described in Section~\ref{sec:Analysis-transformation-synthesis}. It involves reconstructing sequences (B) using one of the latent variables (among $\{\w, \zav, \za, \zv\})$ taken from the sequence (A) and comparing audio and visual attributes extracted from the output sequences to the same attributes extracted from the original sequence (A). This comparison is done using the mean absolute error (MAE) and the Pearson correlation coefficient (PCC). If the MAE metric (resp. the PCC metric) is low (resp. high) for the swapping of a given latent variable, it indicates that the attribute was transferred from the sequence (A) to sequences (B); the swapped variable thus encodes the attribute. For the visual modality, the attributes being considered include the action units (ranging from 0 (not activated) to 1 (very activated)), the angle of the gaze, and the head pose. These factors are estimated using Py-Feat \citep{muhammod2019pyfeat} and Openface \citep{baltruvsaitis2016openface}. For the audio modality, we consider the first two formant frequencies (in Hz) and the pitch (in Hz), estimated using Praat \citep{boersma2021praat} and CREPE \citep{kim2018crepe}. Note that all these attributes are time-varying. The PCC is computed after centering the data (by subtracting the time average of the factor), which is not the case for the MAE. Therefore, contrary to the MAE, the PCC will not be affected by a time-invariant shift of the attribute.

\begin{figure}[t]
     \centering
        \begin{subfigure}[b]{0.23\textwidth}
            \centering
            \includegraphics[width=0.9\textwidth]{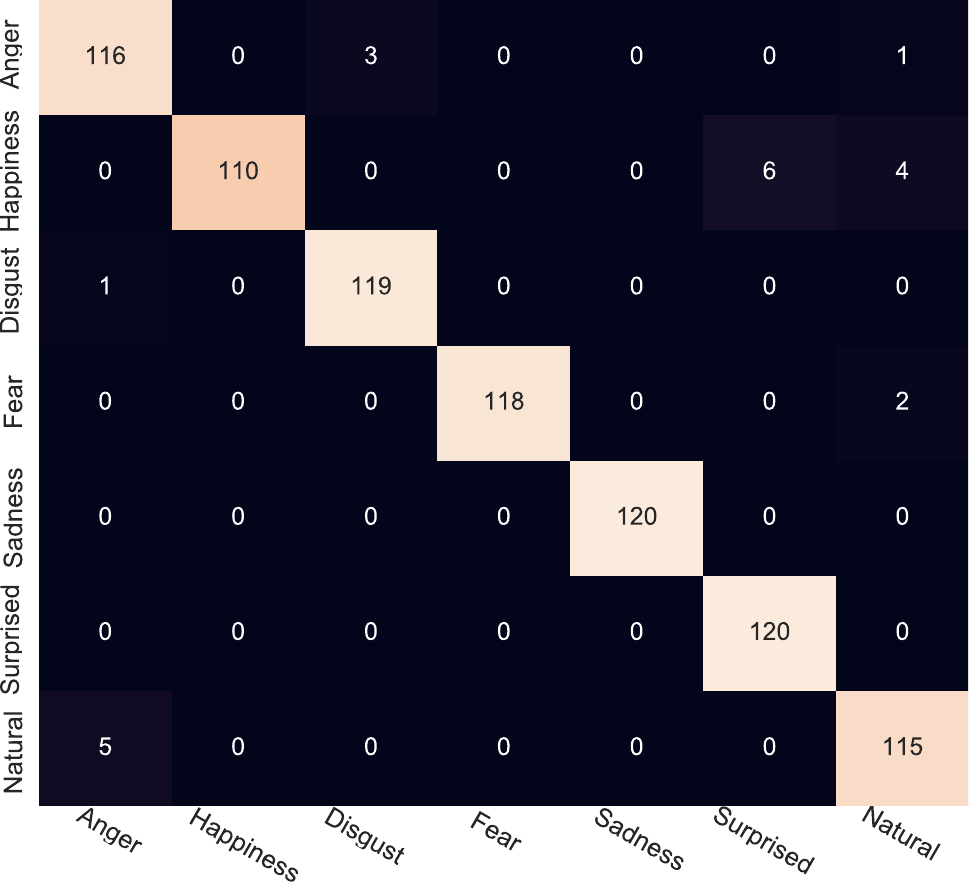}
            \caption{Confusion matrix for emotion classification on the VQ-MDVAE output images after perturbation of the dynamical latent variables.}
            \label{fig:emotion-part1}
        \end{subfigure}
     \hfill
     \begin{subfigure}[b]{0.23\textwidth}
         \centering
            \includegraphics[width=\textwidth]{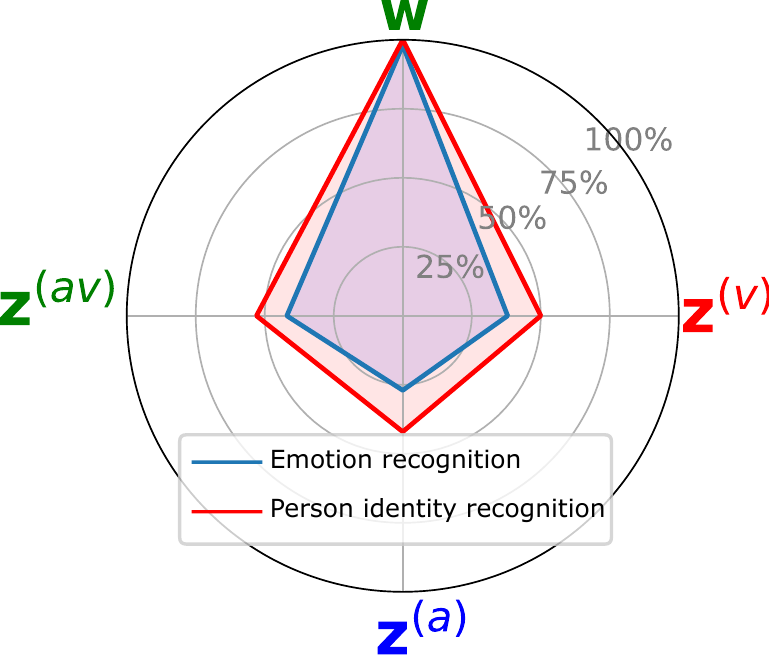}
            \caption{Performance of emotion and person identity recognition for each latent variable of the VQ-MDVAE model.}
            \label{fig:emotion-identification-part1}
     \end{subfigure}
    \hfill
        \caption{Analysis of the latent variables of the VQ-MDVAE model in terms of emotion and person identity.}
        \label{fig:quantitative-emotion-identifiation}
\end{figure}

\noindent\textbf{Discussion}\hspace{0.20cm}
Figure~\ref{fig:corr-mae} presents the average results obtained by repeating the protocol 50 times, i.e., using 50 different sequences (A).
From this figure, we draw four main conclusions. First, the action units related to the lips and jaw (lip press $AU_{24}$, lip parts $AU_{25}$, jaw drop $AU_{26}$, and lip suction $AU_{28}$) and the first two formant frequencies all show high PCC values and low MAE values when performing transformations with the latent variable $\zav$. It indicates that this audiovisual dynamical latent variable plays a significant role in globally controlling these factors. This is very interesting, considering that the lips and jaw are two important speech articulators whose movement induces variations of the shape of the vocal tract and thus also variations of the formant frequencies (the resonance frequencies of the vocal tract). The VQ-MDVAE model thus managed to encode highly-correlated visual and audio factors in the same audiovisual dynamical latent variable.
Secondly, the pitch factor shows a high PCC value when manipulating the dynamical audiovisual latent variable $\zav$. However, it shows a low MAE value when manipulating the static audiovisual latent variable $\w$. This indicates that $\w$ encodes the average pitch value while $\zav$ captures the temporal variation of the pitch around this center value (we remind that the PCC is computed from centered data but not the MAE). The fluctuations in pitch around the average value are encoded in the audiovisual latent variable $\zav$ rather than the audio-specific one $\za$. This finding is supported by a recent study \citep{berry2022correlated} that demonstrates a significant correlation between pitch and the lowering of the jaw. 
Then, the action unit associated with the closing of the eyes ($AU_{43}$), the angle of the gaze as well as the pose of the head show a high PCC and low MAE when manipulating the visual dynamical latent variable $\zv$. This suggests that $\zv$ plays a significant role in globally controlling the movement of the eyelids, the gaze, and the head movements. These factors are indeed much less correlated with the audio than the lip and jaw movements, which explains why they are encoded in the visual dynamical latent variable $\zv$ and not in the audiovisual dynamical latent variable $\zav$.
Finally, action units such as the inner brow raiser ($AU_{01}$), outer brow raiser ($AU_{02}$), upper lid raiser ($AU_{05}$), cheek raiser ($AU_{06}$), and lid tightener ($AU_{07}$) on one side, and nose wrinkler ($AU_{09}$), nasolabial deepener ($AU_{11}$), lip corner puller ($AU_{12}$), and dimpler ($AU_{14}$) on the other side, show high PCC values with respect to $\zv$ and $\zav$, respectively, but low MAE values with respect to $\w$. We argue that this result is related to the encoding of the speaker's emotional state in the latent space of the VQ-MDVAE model. Indeed, we have shown qualitatively that the static audiovisual latent variable $\w$ encodes the global emotional state of a speaker, which explains why it also encodes the average activation level (as indicated by the low MAE values) of the above-mentioned action units that are important for emotions. In contrast, the dynamical latent variables $\zav$ and $\zv$ capture the temporal variations around this average value (as indicated by the high PCC values). As an illustration, we can think of an audiovisual speech utterance spoken by a happy speaker. The global emotional state (happy) would be encoded in $\w$, leading to high constant average values of the cheek raiser ($AU_{06}$) and lip corner puller ($AU_{12}$) action units, and these values would be modulated temporally by the movement of the speech articulators, as encoded in $\zav$.

In Section~\ref{qualitative-parts}, we showed qualitatively that the static audiovisual latent variable $\w$ encodes the speaker's identity and global emotion. 
This paragraph aims to quantify this with two complementary approaches. The first approach operates in the reconstructed image space at the output of the VQ-MDVAE model, while the second approach operates in the latent space of the model. To investigate the emotions in the VQ-MDVAE output images, we randomly select an audiovisual sequence (A) from the test data that is labeled with a specific emotion. We then perturb the dynamical latent variables of (A) by replacing them with those of sequences (B) whose emotions are different from that of sequence (A), while keeping the static audiovisual latent variable $\w$ of sequence (A) unchanged. We evaluate the performance of an emotion classification model (ResMaskNet \citep{pham2021facial}) on the VQ-MDVAE output images produced by this experiment and repeat the process 120 times for each emotion. The results are summarized in a confusion matrix shown in Figure~\ref{fig:emotion-part1}. This matrix is mainly diagonal, indicating that as long as the static audiovisual latent variable $\w$ is not changed, the overall emotion is not changed. This is consistent with the discussion in the previous paragraph, where $\w$ was shown to control the average value of certain action units. In the second approach, we use the latent variable of the VQ-MDVAE model to recognize emotions and identities using a Support Vector Machine (SVM) classifier. The training and test datasets for the SVM comprised 70\% and 30\% of the combined test and validation data from the MEAD dataset, respectively. The dataset consisted of 11 speakers and included eight emotions. The performance accuracy for both classification tasks is shown in Figure~\ref{fig:emotion-identification-part1}, for different latent variables used as input to the classifier. The results show that emotional and identity information are encoded in the static audiovisual latent variable $\w$, with 98\% and 100\% correct classification, respectively. \ref{appendix:visualization} contains visualizations of the static latent space, while the aforementioned companion website provides qualitative results of interpolations on $\w$, demonstrating how we can modify the emotion within an audiovisual speech sequence without altering the identity, and vice versa.

\subsection{Audiovisual facial image denoising}
\label{section:noise-robusstness}

\begin{figure*}[ht]
     \centering
     \begin{subfigure}[b]{0.48\textwidth}
         \centering
         \includegraphics[width=\textwidth]{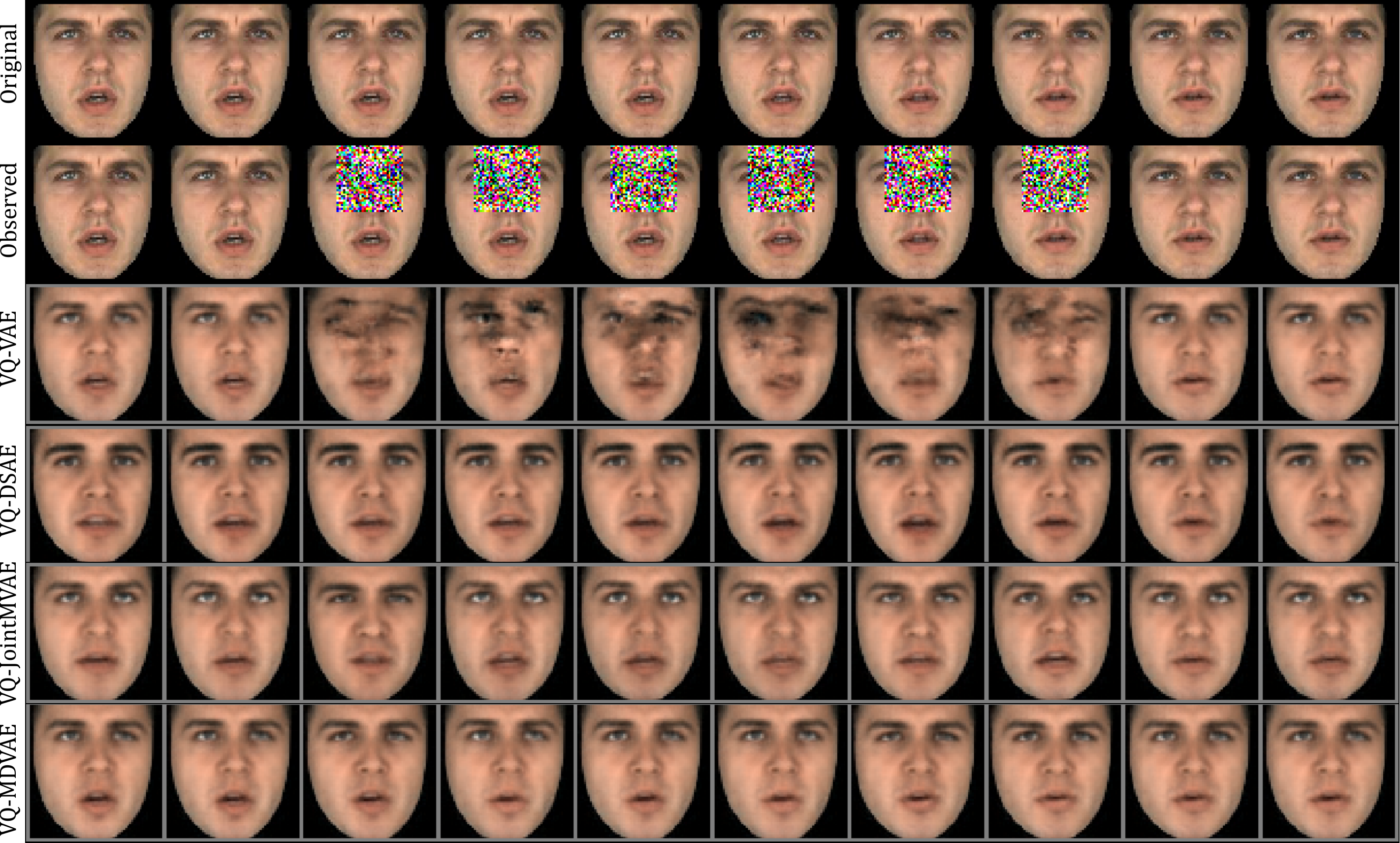}
         \caption{Corruption of the eyes region.}
     \end{subfigure}
     \hfill
     \begin{subfigure}[b]{0.48\textwidth}
         \centering
         \includegraphics[width=\textwidth]{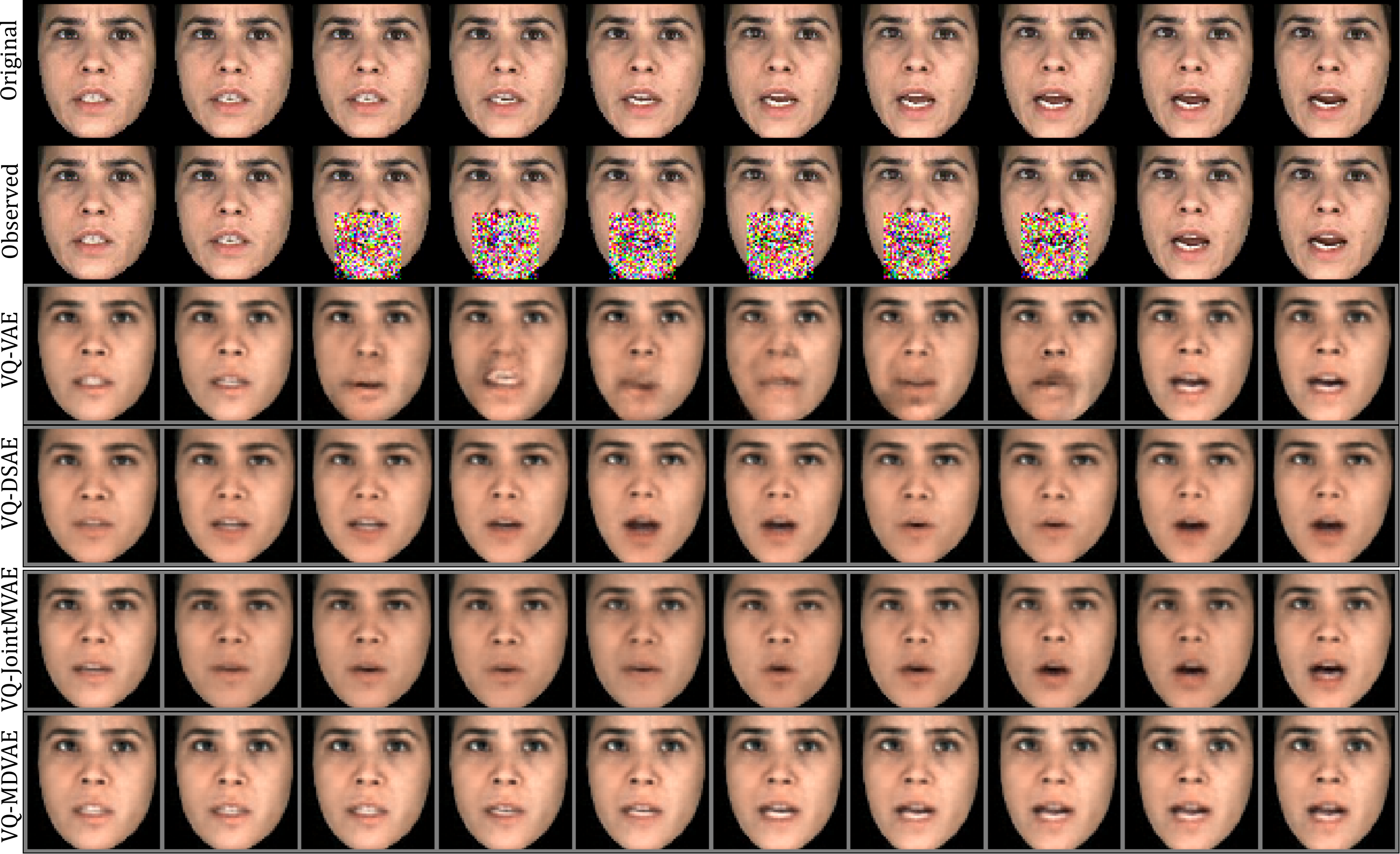}
         \caption{Corruption of the mouth region.}
     \end{subfigure}
    \hfill
     \hfill
        \caption{Qualitative comparison of the denoising results. From top to bottom: perturbed sequences; sequences reconstructed with VQ-VAE; sequences reconstructed with DSAE; sequences reconstructed with VQ-JointMVAE; and sequences reconstructed with VQ-MDVAE.}
        \label{fig:perturbation-qualitative}
\end{figure*}
\begin{figure}[h!]
     \centering
        \begin{subfigure}[b]{0.5\textwidth}
            \centering
            \includegraphics[width=\textwidth]{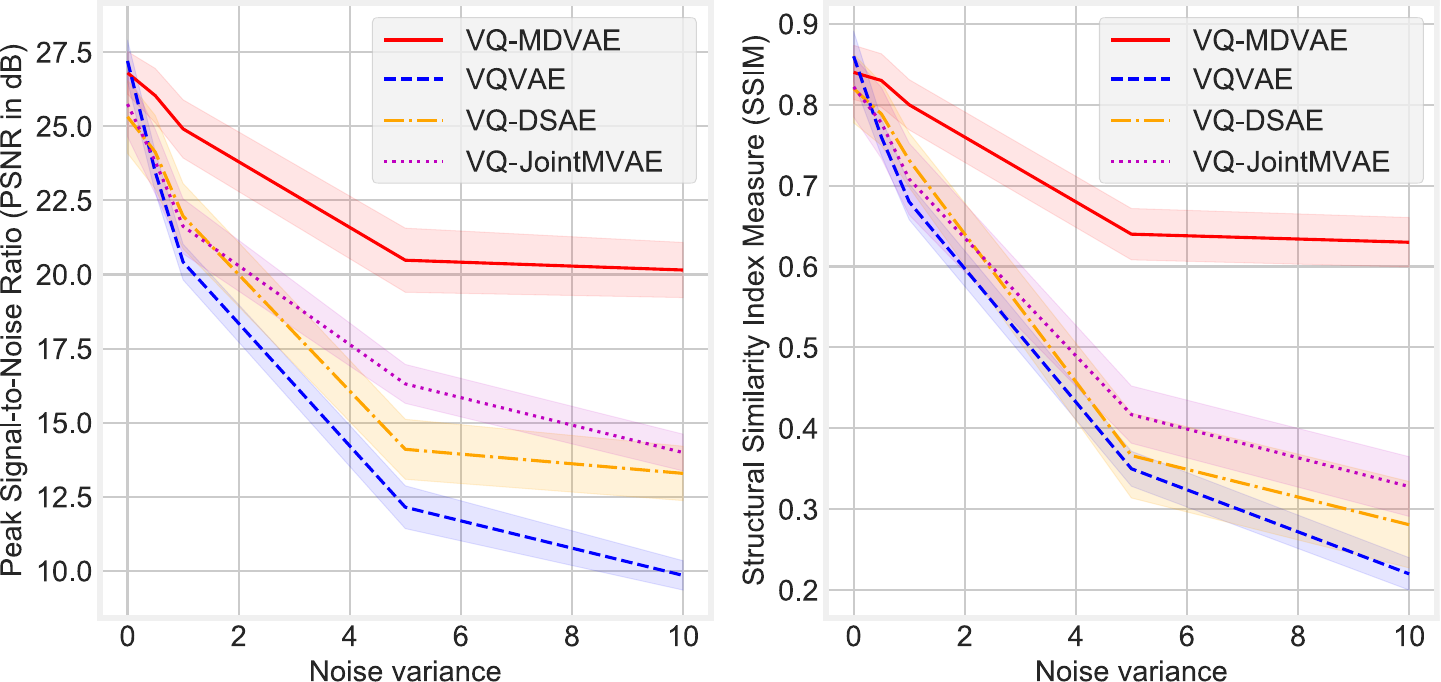}
            \caption{Corruption of the mouth region.}
            \label{fig:mouth_perturbation}
        \end{subfigure}
     \hfill \vspace{0.1cm}
     \begin{subfigure}[b]{0.5\textwidth}
         \centering
            \includegraphics[width=\textwidth]{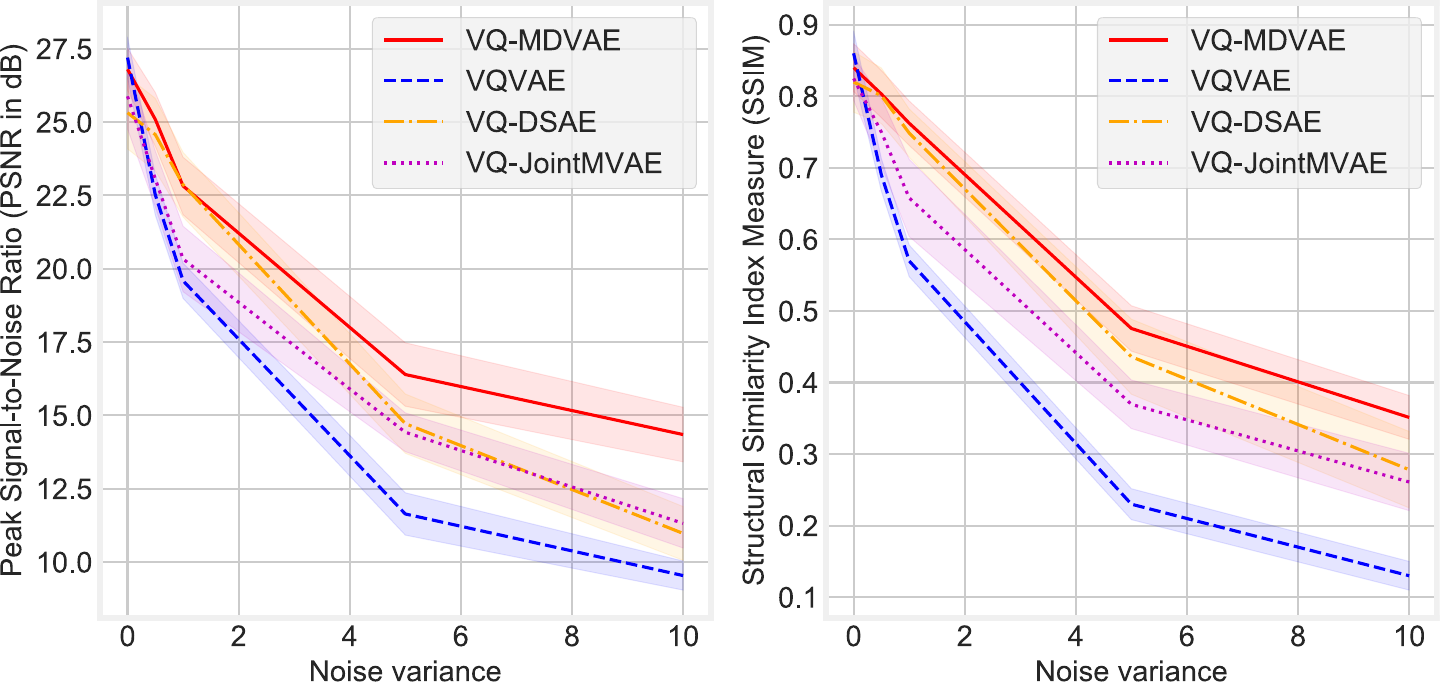}
            \caption{Corruption of the eyes region.}
            \label{fig:eyes_perturbation}
     \end{subfigure}
        \caption{(For better visibility, please zoom in.) Quantitative results of audiovisual facial image denoising. (a) PSNR (left) and SSIM (right) are plotted as a function of the noise variance when the noise is applied to the mouth region. (b) PSNR (left) and SSIM (right) are plotted as a function of the noise variance when the noise is applied to the eyes region.}
        \label{fig:perturbation}
\end{figure}

\noindent\textbf{Experimental set-up}\hspace{0.2cm} This section focuses on denoising audiovisual facial videos. The denoising approach consists of encoding and decoding corrupted visual speech sequences with autoencoder-based models (see next paragraph) pre-trained on the clean MEAD dataset. We intentionally introduced two types of perturbations, strategically located around the eyes and mouth. Specifically, we chose to perturb the sequences using centered isotropic Gaussian noise, and we studied the impact of different levels of noise variance. Our analysis was performed on sequences consisting of ten images, where only the six central images were corrupted.

\noindent\textbf{Methods}\hspace{0.2cm} In this experiment, we compare the performance of VQ-MDVAE, which uses both audio and visual modalities and includes a hierarchical temporal model, with three other models: VQ-VAE~\citep{van2017neural}, a unimodal model only trained on the visual modality and without temporal modeling; DSAE~\citep{li2018disentangled}, a unimodal model only trained on the visual modality and with the same temporal hierarchical model as the proposed VQ-MDVAE; and JointMVAE~\citep{suzuki2016joint}, a multimodal model without temporal modeling.
To ensure a fair comparison, we trained the DSAE and JointMVAE models in two stages, similar to the VQ-MDVAE model. It is important to mention that the VQ-VAE used in this experiment is identical to the one used in VQ-MDVAE, VQ-DSAE, and VQ-JointMVAE.

\noindent\textbf{Metrics}\hspace{0.2cm} To evaluate the denoising performance, we consider again the PSNR and SSIM metrics. These are calculated on the corrupted region of the image, and provide a quantitative measure of the quality and similarity of the denoised image compared to the original. The higher the PSNR and SSIM values, the better the denoising performance.

\noindent\textbf{Discussion}\hspace{0.2cm} 
Figures~\ref{fig:perturbation-qualitative} and~\ref{fig:perturbation} present the qualitative and quantitative results for the denoising experiment, respectively. The mean and standard deviation of the metrics computed over 200 test sequences for the mouth and eyes corruptions are shown in Figures~\ref{fig:mouth_perturbation} and~\ref{fig:eyes_perturbation}, respectively.
Overall, the VQ-MDVAE, VQ-JointMVAE, and VQ-DSAE models outperform the VQ-VAE for both types of perturbations. In the case of mouth corruption, the VQ-MDVAE and VQ-JointMVAE models perform better than the unimodal VQ-DSAE model, demonstrating the benefit of multimodal modeling. These models use the audio modality to denoise the mouth, resulting in a notable 7~dB increase in PSNR at a variance of 10 for VQ-MDVAE compared to VQ-DSAE. As expected, the audio modality is less useful for denoising the eyes, resulting in a smaller advantage for multimodal models in this case. In fact, the PSNR improvement with VQ-MDVAE for the corruption of the eyes is only 3~dB compared to the unimodal VQ-DSAE model. It can also be seen that VQ-MDVAE consistently outperforms VQ-JointMVAE, which shows the benefit of temporal modeling in multimodal models.

\subsection{Audiovisual speech emotion recognition}
This section presents emotion recognition experiments based on the static audiovisual representation $\w$ learned by VQ-MDVAE in an unsupervised manner. We consider two problems: estimating the emotion category and the emotional intensity level.

\noindent\textbf{Experimental set-up}\hspace{0.2cm} We assess the effectiveness of the proposed model on two different datasets: MEAD \citep{wang2020mead} and RAVDESS \citep{livingstone2018ryerson}. The MEAD dataset was presented in Section~\ref{sec:audiovisual-dataset}. The RAVDESS dataset contains 1440 audio files that were recorded by 24 professional actors, with each file labeled with one of eight different emotions: neutral, calm, happy, sad, angry, fearful, disgusted, or surprised.
We conduct two types of evaluations to measure performance. 
The first evaluation involves recognizing emotions and their intensity levels in the case where individuals can be seen during the training phase (\emph{person-dependent evaluation}). For this evaluation, we randomly divide the dataset into 70\% training data and 30\% testing data.
The second evaluation involves recognizing emotions and their intensity levels in the case where individuals are not seen during the training phase (\emph{person-independent evaluation}). To perform this evaluation, we use a 5-fold cross-validation approach to separate the speakers' identities between the training and evaluation sets.
Through these evaluations, we are able to assess the ability of the models to detect emotions and their intensity levels in both person-dependent and person-independent scenarios using two different datasets.

\noindent\textbf{Methods}\hspace{0.2cm} We compare the performance of VQ-MDVAE with several methods from the literature. First, the VQ-DSAE-audio and VQ-DSAE-visual models, which correspond to the VQ-DSAE model already discussed in the previous experiments, here trained either on the audio modality or on the visual modality. We remind that VQ-DSAE is an improved version of DSAE \citep{li2018disentangled} that uses the 2-stage training process proposed in the present paper. VQ-MDVAE can be seen as a multimodal extension of VQ-DSAE because both methods share the same hierarchical temporal model, including a static and a dynamical latent variable. Comparing VQ-MDVAE with the two VQ-DSAE models will thus allow us to fairly assess the benefit of a multimodal approach to emotion recognition. Second, the wav2vec model \citep{schneider2019wav2vec}, which is a self-supervised unimodal representation learning approach. Wav2vec is trained on the audio speech signals of the Librispeech dataset \citep{panayotov2015librispeech}, which includes $960$ hours of unlabeled speech data, with $2 338$ different speakers. Finally, we also include in this experiment two state-of-the-art supervised multimodal approaches \citep{chumachenko2022self, tsai2019multimodal}, which are based on an audiovisual transformer architecture. The method of \citet{chumachenko2022self} will be referred to as ``AV transformer''. It also relies on transfer learning using EfficientFace \citep{zhao2021robust}, a model pre-trained on AffectNet \citep{mollahosseini2017affectnet}, the largest dataset of in-the-wild facial images labeled in emotions. The method of \citet{tsai2019multimodal} will be referred to as ``MULT'' for multimodal transformer.

AV transformer and MULT are fully supervised, trained, and evaluated on RAVDESS. This contrasts with wav2vec, VQ-DSAE-audio, VQ-DSAE-visual, and VQ-MDVAE, which are pre-trained in a self-supervised or unsupervised manner and then used as frozen feature extractors to train a small classification model on top of the extracted representation of (audiovisual) speech.
For VQ-DSAE-audio, VQ-DSAE-visual, and VQ-MDVAE, only the global latent variable ($\w$) is fed to the classifier. For wav2vec, a temporal mean-pooling layer is added before the classifier as in \citep{pepino2021emotion}. Depending on the feature extraction method and evaluation configuration (person independent or dependent), we consider different classification models: a simple multinomial logistic regression (MLR) implemented with a single linear layer followed by a softmax activation function, or a multilayer perceptron (referred to as MLP) with two hidden layers followed by a linear layer and a softmax activation function. In the person-dependent setting, we explore a third approach (referred to as DA + MLR) that involves transforming the test data using an unsupervised domain adaptation method (DA) before classification with the MLR model. Unsupervised domain adaptation is here used to compensate for the domain shift due to the fact that speakers are different in the training and testing sets. This is further discussed below.

\begin{figure}[t]
\centering
\includegraphics[width=\linewidth]{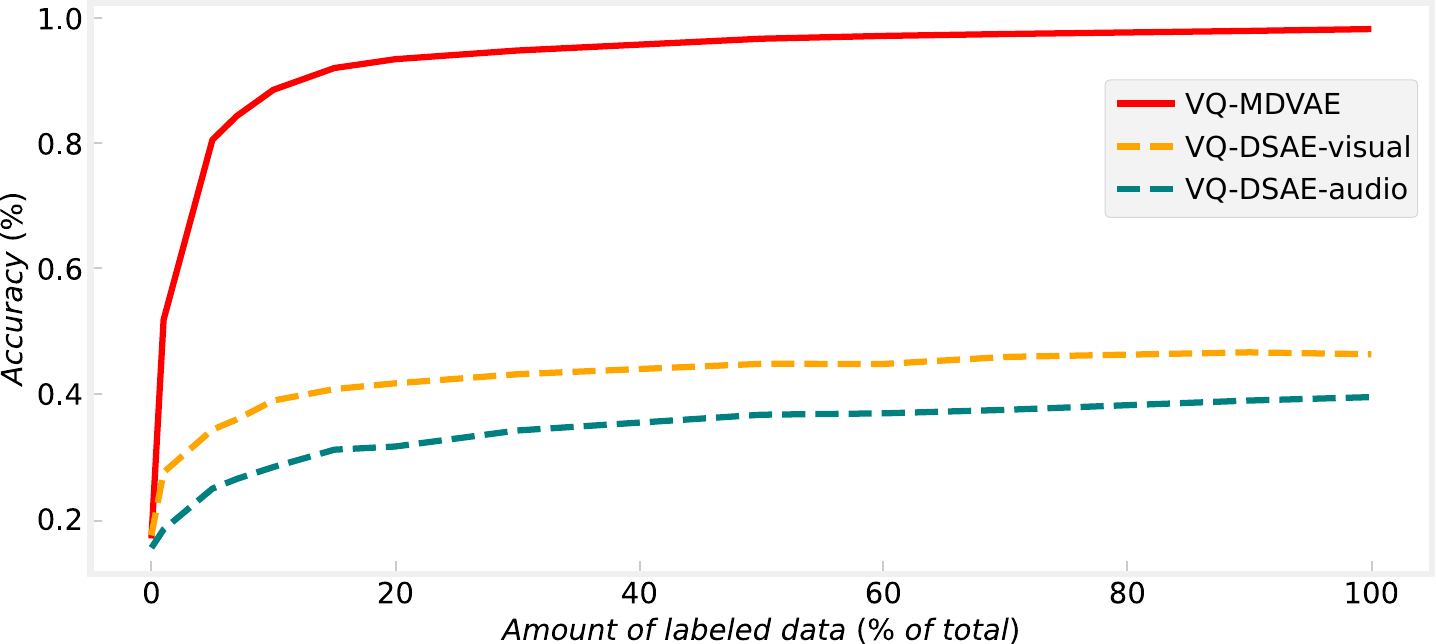}
\caption{Accuracy for emotion category classification as a function of the amount of labeled data used to train the MLR classification model on the MEAD dataset in the person-dependent evaluation setting.}
\label{fig:acc_vs_amout_of_data}
\end{figure}

\noindent\textbf{Discussion}\hspace{0.2cm} We start by comparing VQ-MDVAE with its two unimodal counterparts, VQ-DSAE-audio and VQ-DSAE-visual, for the emotion category classification task on the MEAD dataset. In Figure~\ref{fig:acc_vs_amout_of_data}, we show the classification accuracy as a function of the amount of labeled training data used to train the MLR classification model. VQ-MDVAE and the VQ-DSAE models are all pre-trained in an unsupervised manner on the MEAD dataset. Using the exact same experimental protocol, we observe that when using 100\% of the labeled data the VQ-MDVAE model outperforms its two unimodal counterparts by about 50\% of accuracy, which clearly demonstrates the interest of a multimodal approach to emotion recognition from latent representations learned with dynamical VAEs. Another interesting observation is that we need less than 10\% of the labeled data to reach 90\% of the maximal performance of the VQ-MDVAE model. 

\begin{figure*}[t]
\begin{minipage}{1\textwidth}
\vspace{0.2cm}
\captionof{table}{Accuracy (\%) and F1-score (\%) results of emotion category and intensity level recognition in the person-dependent (PD) and person-independent (PI) evaluation settings for the MEAD and RAVDESS datasets. The best scores are in bold and second best scores are underlined. For the VQ-MDVAE model evaluated on RAVDESS, two scores are reported. The first one corresponds to VQ-MDVAE trained on MEAD only, and the second one to the same model fine-tuned (in an unsupervised manner) on RAVDESS.}
\label{tab:emo_recog_results}
\resizebox{1.0\linewidth}{!}{ 
{\renewcommand{\arraystretch}{1.2}
\begin{tabular}{ccccccccccc}
\hline \hline
& \multicolumn{2}{c}{\multirow{2}{*}{Model}} & \multicolumn{4}{c}{Emotion category} & \multicolumn{4}{c}{Emotion intensity level} \\
& \multicolumn{2}{c}{} & \multicolumn{2}{c}{MEAD} & \multicolumn{2}{c}{RAVDESS} & \multicolumn{2}{c}{MEAD} & \multicolumn{2}{c}{RAVDESS} \\
& classification & representation  & Accuracy & F1-score & Accuracy & F1-score & Accuracy & F1-score & Accuracy & F1-score \\ \hline\hline
\multirow{5}{*}{PD} & \multirow{4}{*}{MLR} & VQ-DSAE-audio \citep{li2018disentangled} & 40.4 & 39.3 & - & - & 48.7 & 45.7 & - & - \\
& & VQ-DSAE-visual \citep{li2018disentangled} & 50.6 & 51.1 & - & - & 43.3 & 44.2 & - & - \\
& & wav2vec \citep{schneider2019wav2vec} & {\ul 76.2} & {\ul 75.0} & 74.3 & 75.5 & {\ul 55.0} & {\ul 54.6} & 76.5 & 76.3 \\
& & VQ-MDVAE (our) & \textbf{98.2} & \textbf{98.3} & 81.9 / \textbf{89.4} & 82.9 / \textbf{89.6} & \textbf{83.9} & \textbf{83.1} & {\ul 78.0} / \textbf{80.1} & {\ul 77.2} / \textbf{79.8} \\ 
\cmidrule(lr){2-3}
& \multicolumn{2}{c}{AV transformer \citep{chumachenko2022self}} & - & - & {\ul 89.2} & {\ul 88.6} & - & - & - & - \\[.2cm]
 \hline \hline
\multirow{6}{*}{PI} & \multirow{2}{*}{MLR} & wav2vec \citep{schneider2019wav2vec} & 68.4 & 64.5 & 69.5 & 68.6 & 51.8 & 50.3 & 76.6 & 75.6 \\
 & & VQ-MDVAE (our) & 73.2 & 72.5 & 68.8 / 71.4 & 68.5 / 70.5 & 63.8 & 61.7 & 73.8 / 77.2 & 75.7 / 77.6 \\ \cmidrule(lr){2-3}
& \multirow{2}{*}{MLP} & wav2vec \citep{schneider2019wav2vec} & 70.9 & 70.8 & 70.2 & 70.6 & 53.7 & 53.9 & 76.6 & 76.3 \\ 
& & VQ-MDVAE (our) & {\ul 80.0} & {\ul 80.5} & 77.5 / 78.7 & 78.0 / 78.1 & {\ul 71.5} & {\ul 72.2} & 77.4 / 77.4 & 77.6 / 77.7 \\ \cmidrule(lr){2-3}
& \multirow{2}{*}{DA + MLR} & wav2vec \citep{schneider2019wav2vec} & 71.0 & 69.9 & 71.6 & 71.2 & 53.5 & 52.9 & 76.8 &  76.5 \\
 & & VQ-MDVAE (our) & \textbf{83.1} & \textbf{82.2} & 78.1 / \textbf{79.3} & 78.0 / \textbf{80.7} & \textbf{77.5} & \textbf{78.0} & {\ul 78.1} / \textbf{79.0} & {\ul 78.5} / \textbf{79.1} \\\cmidrule(lr){2-3}
& \multicolumn{2}{c}{MULT \citep{tsai2019multimodal}} & - & - &  76.6 &  77.3 & - & - & - & - \\
& \multicolumn{2}{c}{AV transformer \citep{chumachenko2022self}} & - & - & {\ul 79.2} & {\ul 78.2} & - & - & - & - \\[.2cm]
\hline \hline
\end{tabular}
}
}
\end{minipage}
\end{figure*}

Table~\ref{tab:emo_recog_results} compares the emotion category and intensity level classification performance of the proposed VQ-MDVAE method and the previously mentioned methods from the literature. We report the accuracy (in \%), defined as the ratio of correctly predicted instances to the total number of instances, and the F1-score (in \%), defined as the harmonic mean of the precision and recall.
For the person-dependent evaluation (``PD'' section of the table), VQ-MDVAE demonstrates superior performance in recognizing emotion categories (resp. emotion levels) on the MEAD dataset, outperforming VQ-DSAE-audio by 57.8\% (resp. 35.2\%), VQ-DSAE-visual by 47.6\% (resp. 40.6\%), and wav2vec by 22\% (resp. 28.5\%) of accuracy. On the RAVDESS dataset, it can be observed that VQ-MDVAE pre-trained on MEAD and finetuned on RAVDESS (in an unsupervised manner) outperforms the fully-supervised state-of-the-art method \citep{chumachenko2022self} (AV transformer) by 0.2\% of accuracy and 1.0\% of F1-score. Note that the AV transformer cannot be trained simultaneously on MEAD and RAVDESS because the emotion labels in these two datasets are different. On the contrary, the proposed VQ-MDVAE model can be pre-trained on any emotional audiovisual speech dataset, precisely because it is unsupervised. The learned representation can then be used to train a supervised classification model. 
This evaluation confirms that the static audiovisual latent variable $\w$ learned by the proposed VQ-MDVAE is an effective representation for audiovisual speech emotion recognition. Indeed, as shown in Figure \ref{fig:2D-visualisation} of \ref{appendix:visualization}, emotion categories and levels form distinct clusters in the static audiovisual latent space of the VQ-MDVAE model.

For the person-independent evaluation, we only compare VQ-MDVAE, wav2vec, MULT and AV transformer, as the person-dependent evaluation showed that VQ-MDVAE outperforms its two unimodal counterparts based on VQ-DSAE. Compared with the person-dependent setting, we observe in the ``PI'' section of Table~\ref{tab:emo_recog_results} a decrease in performance for all methods using an MLR classification model. For VQ-MDVAE, this decline can be analyzed through visual representations of the static audiovisual latent space, as shown in Figure~\ref{fig:2D-visualisation} of \ref{appendix:visualization}. This figure highlights the hierarchical structure of the static latent audiovisual space in terms of identity, emotion, and intensity level. In this structure, identities are represented by clusters, each of which is made up of several emotion clusters. These clusters represent eight distinct emotions distributed in a range of intensity levels from weak to strong. 
As a result, each identity is associated with its own representation of emotions, which means that the emotion clusters differ from one identity to another. By incorporating the identity information as in the previous evaluation approach, we can more accurately classify the emotion categories. Consequently, a simple linear model (MLR) is sufficient for classifying both the emotions and their levels. 
To improve generalization to test data where speakers were not seen during training, we propose two solutions.  
First, we improve the classification model by replacing the linear MLR classifier by a non-linear MLP classifier, which results in a substantial increase in accuracy for the VQ-MDVAE model: +6.8\% and +7.3\% for emotion category classification on the MEAD and RAVDESS datasets, respectively (using the finetuned model for RAVDESS). We observe a similar trend with the wav2vec + MLP model, which leads to an improvement in performance compared to using the MLR classifier.
Second, we keep the MLR classification model but apply unsupervised domain adaptation to the test data using an optimal transport approach \citep{courty2017joint}. Domain adaptation has been shown to be effective when dealing with domain shifts caused by unknown transformations, such as changes in identity, gender, age, ethnicity, or other factors \citep{wei2018unsupervised, kim2022optimal}. 
To adapt our model to a new domain, we use optimal transport to map the probability distribution of the source domain ($\w$ of seen identities) to that of the target domain ($\w$ of unseen identities). This is accomplished by minimizing the earth mover's distance between the two distributions \citep{courty2017joint}. By finding an optimal transport plan, we can transfer knowledge from the source domain to the target domain in an unsupervised manner (i.e., emotion labels are not used), resulting in a large improvement in accuracy for both the wav2vec and VQ-MDVAE models compared to when no domain adaptation is performed: +9.9\% and +7.9\% for emotion category classification on the MEAD and RAVDESS datasets with the VQ-MDVAE model (using the finetuned model for RAVDESS), and +2.6\% and +2.1\% with the wav2vec model. It can also be seen that the MLR linear classification model with domain adaptation is more effective than the MLP non-linear classification approach. Finally, for emotion category classification on RAVDESS, we see that the proposed VQ-MDVAE (finetuned) with domain adaptation and MLR outperforms the state-of-the-art fully-supervised AV transformer and MULT methods by 0.1\%  and 2.7\% of accuracy. This is particularly interesting considering that most of the proposed model parameters have been learned in an unsupervised manner. Indeed, only the MLR classification model, which includes 680 ($84 \times 8 + 8$) trainable parameters, is learned using labeled emotional audiovisual speech data.

\section{Conclusion}
\label{section:conclusion}
Deep generative modeling is a powerful unsupervised learning paradigm that can be applied to many different types of data. In this paper, we proposed the VQ-MDVAE model to learn structured and interpretable representations of multimodal sequential data. A key to learn a meaningful representation in the proposed approach is to structure the latent space into different latent variables that disentangle static, dynamical, modality-specific and modality-common information. By defining appropriate probabilistic dependencies between the observed data and the latent variables, we were able to learn structured and interpretable representations in an unsupervised manner. Trained on an expressive audiovisual speech dataset, the same VQ-MDVAE model was used to address several tasks in audiovisual speech processing. This versatility contrasts with task-specific supervised models. The experiments have shown that the VQ-MDVAE model effectively combines the audio and visual information in static ($\w$) and dynamical ($\zav$) audiovisual latent variables, while characteristics specific to each individual modality are encoded in dynamical modality-specific latent variables ($\za$ and $\zv$). Indeed, we have shown that lip and jaw movements can be synthesized by transferring $\zav$ from one sequence to another, while preserving the speaker's identity, emotional state, and visual-only facial movements. For denoising, we have shown that the audio modality provides robustness with respect to the corruption of the visual modality on the mouth region. Finally, we proposed to use the static audiovisual latent variable $\w$ for emotion recognition. This approach was shown to be effective with only a few labeled data, and it obtained much better accuracy than unimodal baselines. Experimental results have also shown that the proposed unsupervised representation learning approach outperforms state-of-the-art fully-supervised emotion recognition methods based on an audiovisual transformer.

Unfortunately, the two modalities are not always available in audiovisual speech processing. For instance, the audio modality might be missing due to highly intrusive noise, and the visual modality might be missing due to low-lighting conditions. A robust multimodal information retrieval system should be able to handle such a situation where some modalities are temporarily missing. In the current configuration of the MDVAE model, the proposed approach relies on both modalities for inference of the latent variables. Nevertheless, MDVAE could be extended to accommodate single-modality inference using the ``sub-sampled training'' approach proposed in \citep{wu2018multimodal}, or maybe using the multimodal masking strategies proposed in \citep{bachmann2022multimae}. Moreover, being able to infer all latent variables from one single modality would allow the model to be used for cross-modality generation, i.e., generating one modality given another.

\appendix
\section{The detailed architecture of the vector quantized MDVAE}
\label{appendix:architecture}
This section details the architecture of the VQ-MDVAE model, starting with the VQ-VAE and then the MDVAE. 

\subsection{VQ-VAE}

The VQ-VAE developed for audio or images consists of three parts: (i) an encoder that maps an image to a sequence of continuous latent variables, referred to as the intermediate representation in the paper, (ii) a shared codebook that is used to quantize these continuous latent vectors to a set of discrete latent variables (each vector is replaced with the nearest vector from the codebook), and (iii) a decoder that maps the indices of the vectors from the codebook back to an image. The architectures of the visual and audio VQ-VAEs are described in tables~\ref{tab:visual-VQ-VAE-architecture} and~\ref{tab:audio-VQ-VAE-architecture}, respectively.

\subsection{MDVAE}
\begin{figure*}[]
    \centering
    \includegraphics[width=\textwidth]{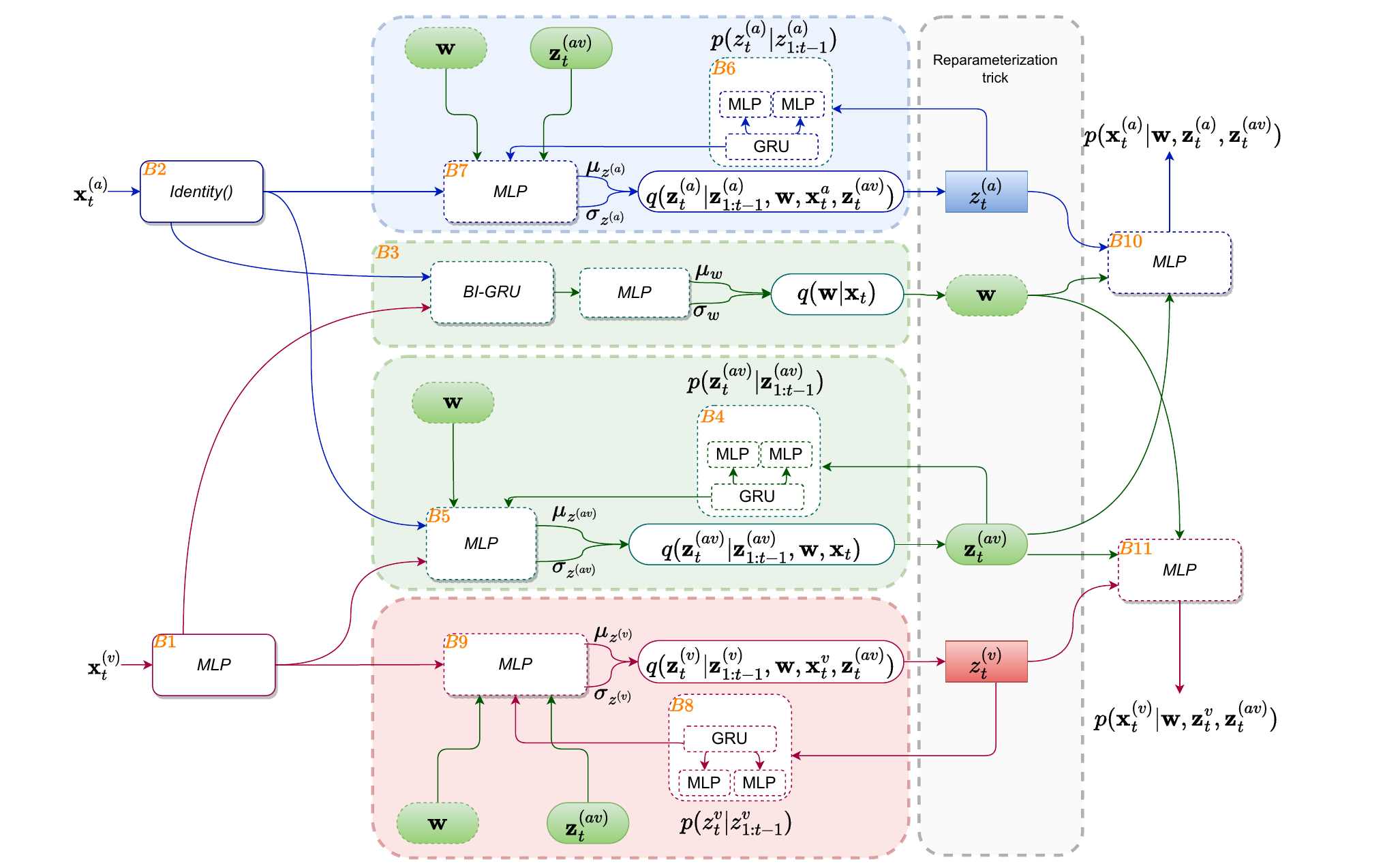}
    \caption{(Better zoom in) The overall architecture of the MDVAE.}
    \label{fig:architecture-MDVAE}
\end{figure*}

MDVAE is decomposed into two models: (i) the first is the inference model (encoder), which is further decomposed into four inferences for each latent variable, represented by Gaussian distributions whose parameters are determined via a neural network. The prior distributions for the dynamic latent variables are also trained, except for the static latent space, where the prior is assumed to be a standard normal distribution. (ii) The second part is composed of two decoders, one for the visual modality and the other for the audio modality. Structured only with linear layers and non-linear activation functions, the input of these two decoders are the concatenation of ${\w, \zavt, \zvt}$ and ${\w, \zavt, \zat}$ for the visual and audio modalities, respectively. Table~\ref{tab:details-architecture} and Figure~\ref{fig:architecture-MDVAE} present the details of the MDVAE architecture. The figure provides an overview of the MDVAE architecture, including the connections between the blocks and the variables. The table complements the figure by detailing each block individually, including its dimensions, activation functions, and other relevant information. Together, the table and figure provide a comprehensive description of the MDVAE architecture.

\begin{table}[]
\caption{The architecture of the VQ-VAE-visual.}
\label{tab:visual-VQ-VAE-architecture}
\resizebox{1.0\linewidth}{!}{ 
{\renewcommand{\arraystretch}{1.2}
\begin{tabular}{cccc}
\hline \hline
 & Layer  & Activation & Output dim \\ \hline\hline
Input & - & - & \texttt{3 $\times$ 64 $\times$ 64} \\
\cmidrule(lr){2-3}
\multirow{5}{*}{Encoder} & \texttt{Conv2D(3, 64, 4, 2, 1)} & ReLu & \texttt{64 $\times$ 32 $\times$ 32} \\
& \texttt{Conv2D(64, 128, 4, 2, 1)} & ReLu & \texttt{128 $\times$ 16 $\times$ 16} \\
& \texttt{Conv2D(128, 128, 4, 2, 1)} & ReLu & \texttt{128 $\times$ 8 $\times$ 8} \\
& \texttt{2 $\times$ Residual Stack}   & ReLu & \texttt{128 $\times$ 8 $\times$ 8} \\
& \texttt{Conv2D(128, 32, 1, 1)} & - & \texttt{32 $\times$ 8 $\times$ 8} \\ \cmidrule(lr){2-3}
\multirow{5}{*}{Decoder} & \texttt{ConvT2D(32, 128, 1, 1)} & - & \texttt{128 $\times$ 8 $\times$ 8} \\
& \texttt{2 $\times$ Residual Stack (T)} & ReLu & \texttt{128 $\times$ 8 $\times$ 8} \\
& \texttt{ConvT2D(128, 64, 4, 2, 1)} & ReLu & \texttt{128 $\times$ 16 $\times$ 16} \\
& \texttt{ConvT2D(64, 64, 4, 2, 1)} & ReLu & \texttt{64 $\times$ 32 $\times$ 32} \\
& \texttt{ConvT2D(64, 3, 4, 2, 1)} & - & \texttt{3 $\times$ 64 $\times$ 64} \\
\\ \cmidrule(lr){1-4}
\multicolumn{4}{c}{\texttt{Conv2D(in\_channel, out\_channel, kernel\_size, stride, padding)}}\\
\multicolumn{4}{c}{\texttt{Residual Stack (T)} = \{ \texttt{2 $\times$ Conv(T)2D(128, 128, 3, 1, 1)}\}} \\
\hline \hline
\end{tabular}
}
}
\end{table}

\begin{table}[]
\caption{The architecture of the VQ-VAE-audio.}
\label{tab:audio-VQ-VAE-architecture}
\resizebox{1.0\linewidth}{!}{ 
{\renewcommand{\arraystretch}{1.2}
\begin{tabular}{cccc}
\hline \hline
& Layer  & Activation & Output dim \\ \hline\hline
Input & - & - & \texttt{1 $\times$ 513} \\
\cmidrule(lr){2-3}
\multirow{5}{*}{Encoder} & \texttt{Conv1D(1, 16, 4, 2, 1)} & Tanh & \texttt{16 $\times$ 256} \\
& \texttt{Conv1D(16, 32, 4, 2, 1)} & Tanh & \texttt{32 $\times$ 128} \\
& \texttt{Conv1D(32, 32, 3, 2, 1)} & Tanh & \texttt{32 $\times$ 64} \\
& \texttt{1 $\times$ Residual Stack}   & Tanh & \texttt{32 $\times$ 64} \\
& \texttt{Conv1D(32, 8, 1, 1)} & - & \texttt{8 $\times$ 64} \\ \cmidrule(lr){2-3}
\multirow{5}{*}{Decoder} & \texttt{ConvT1D(8, 32, 1, 1)} & - & \texttt{32 $\times$ 64} \\
& \texttt{1 $\times$ Residual Stack (T)} & Tanh & \texttt{32 $\times$ 64} \\
& \texttt{ConvT1D(32, 32, 3, 2, 1)} & Tanh & \texttt{32 $\times$ 128 } \\
& \texttt{ConvT1D(32, 16, 4, 2, 1)} & Tanh & \texttt{16 $\times$ 256} \\
& \texttt{ConvT1D(16, 1, 4, 2, 0)} & - & \texttt{1 $\times$ 513} \\
\\ \cmidrule(lr){1-4}
\multicolumn{4}{c}{\texttt{Conv1D(in\_channel, out\_channel, kernel\_size, stride, padding)}}\\
\multicolumn{4}{c}{\texttt{Residual Stack (T)} = \{ \texttt{2 $\times$ Conv(T)1D(32, 32, 3, 1, 1)}\}} \\
\hline \hline
\end{tabular}
}
}
\end{table}

\begin{table}[]
\caption{The architecture details of the MDVAE. The blocks from B1 to B11 are illustrated in Figure~\ref{fig:architecture-MDVAE} to better understand their interactions.}
\label{tab:details-architecture}
\resizebox{1.0\linewidth}{!}{ 
{\renewcommand{\arraystretch}{1.2}
\begin{tabular}{cccc}
\hline \hline
Block & Layer  & Activation & Output dim. \\ \hline\hline
\multirow{2}{*}{$B1$} & \texttt{Linear($32 \cdot 8 \cdot 8$, 1024)} & ReLu & \texttt{1024} \\
& \texttt{Linear(1024, 512)} & ReLu & $r_v$ = \texttt{512}
\\ \cmidrule(lr){2-3}
$B2$ & \texttt{Identity} & - & $r_a$ = \texttt{512}
\\ \cmidrule(lr){2-3}
\multirow{4}{*}{$B3$} & \texttt{GRU($r_v + r_a$, 256, 1, True)} & - & 2 $\cdot$ \texttt{256} \\
& \texttt{Linear(2 $\cdot$ 256, 256)} & Tanh & \texttt{256} \\
& $\sigma_{\w}$: \texttt{Linear(256, $l_w$)} & - & \texttt{$l_w$} \\
& $\mu_{\w}$: \texttt{Linear(256, $l_w$)} & - & \texttt{$l_w$} 
\\ \cmidrule(lr){2-3}
\multirow{4}{*}{$B4$} & \texttt{GRU($l_{av}$, 128, 1, False)} & - & $h_{av}$ \\
& \texttt{Linear(128, 64)} & ReLu & \texttt{64} \\
& \texttt{Linear(64, $l_{av}$)} & - & \texttt{$l_{av}$} \\
& \texttt{Linear(64, $l_{av}$)} & - & \texttt{$l_{av}$}
\\ \cmidrule(lr){2-3}
\multirow{4}{*}{$B5$} & \texttt{Linear($r_v + r_a + h_{av} + l_{w}$, 256)} & ReLu & $256$ \\
& \texttt{Linear(256, 128)} & ReLu & \texttt{128} \\
& $\sigma_{\zav}$: \texttt{Linear(128, $l_{av}$)} & - & \texttt{$l_{av}$} \\
& $\mu_{\zav}$: \texttt{Linear(128, $l_{av}$)} & - & \texttt{$l_{av}$}
\\ \cmidrule(lr){2-3}
\multirow{4}{*}{$B6$} & \texttt{GRU($l_{a}$, 128, 1, False)} & - & $h_{a}$ \\
& \texttt{Linear(128, 32)} & ReLu & \texttt{32} \\
& \texttt{Linear(32, $l_{a}$)} & - & \texttt{$l_{a}$} \\
& \texttt{Linear(32, $l_{a}$)} & - & \texttt{$l_{a}$}
\\ \cmidrule(lr){2-3}
\multirow{4}{*}{$B7$} & \texttt{Linear($r_a + h_{a} + l_{w}$, 128)} & Tanh & $128$ \\
& \texttt{Linear(128, 32)} & Tanh & \texttt{32} \\
& $\sigma_{\za}$: \texttt{Linear(32, $l_{a}$)} & - & \texttt{$l_{a}$} \\
& $\mu_{\za}$: \texttt{Linear(32, $l_{a}$)} & - & \texttt{$l_{a}$}
\\ \cmidrule(lr){2-3}
\multirow{4}{*}{$B8$} & \texttt{GRU($l_{v}$, 128, 1, False)} & - & $h_{v}$ \\
& \texttt{Linear(128, 64)} & ReLu & \texttt{64} \\
& \texttt{Linear(64, $l_{v}$)} & - & \texttt{$l_{v}$} \\
& \texttt{Linear(64, $l_{v}$)} & - & \texttt{$l_{v}$}
\\ \cmidrule(lr){2-3}
\multirow{4}{*}{$B9$} & \texttt{Linear($r_v + h_{v} + l_{w}$, 256)} & ReLu & $256$ \\
& \texttt{Linear(256, 128)} & ReLu & \texttt{128} \\
& $\sigma_{\zv}$: \texttt{Linear(128, $l_{v}$)} & - & \texttt{$l_{v}$} \\
& $\mu_{\zv}$: \texttt{Linear(128, $l_{v}$)} & - & \texttt{$l_{v}$}
\\ \cmidrule(lr){2-3}
\multirow{3}{*}{$B{10}$} & \texttt{Linear($l_v + l_{av} + l_w$, 512)} & ReLu & \texttt{512} \\
& \texttt{Linear(512, 1024)} & ReLu & \texttt{1024}\\
& \texttt{Linear(1024, 2048)} & ReLu & \texttt{2048}
\\ \cmidrule(lr){2-3}
\multirow{3}{*}{$B{11}$} & \texttt{Linear($l_a + l_{av} + l_w$, 128)} & Tanh & \texttt{128} \\
& \texttt{Linear(128, 256)} & Tanh & \texttt{256} \\
& \texttt{Linear(256, 512)} & Tanh & \texttt{512} \\
\\ \cmidrule(lr){1-4} 
\multicolumn{4}{c}{\texttt{GRU(input\_size, hidden\_size, num\_layers, bidirectional)}} \\
\multicolumn{4}{c}{\texttt{Linear(input\_size, output\_size)}} \\
\hline \hline
\end{tabular}
}
}
\end{table}

\section{Visualization of the MDVAE static latent space}
\label{appendix:visualization}

\begin{figure}[]
     \centering
        \begin{subfigure}[b]{0.48\textwidth}
            \centering
            \includegraphics[width=\textwidth]{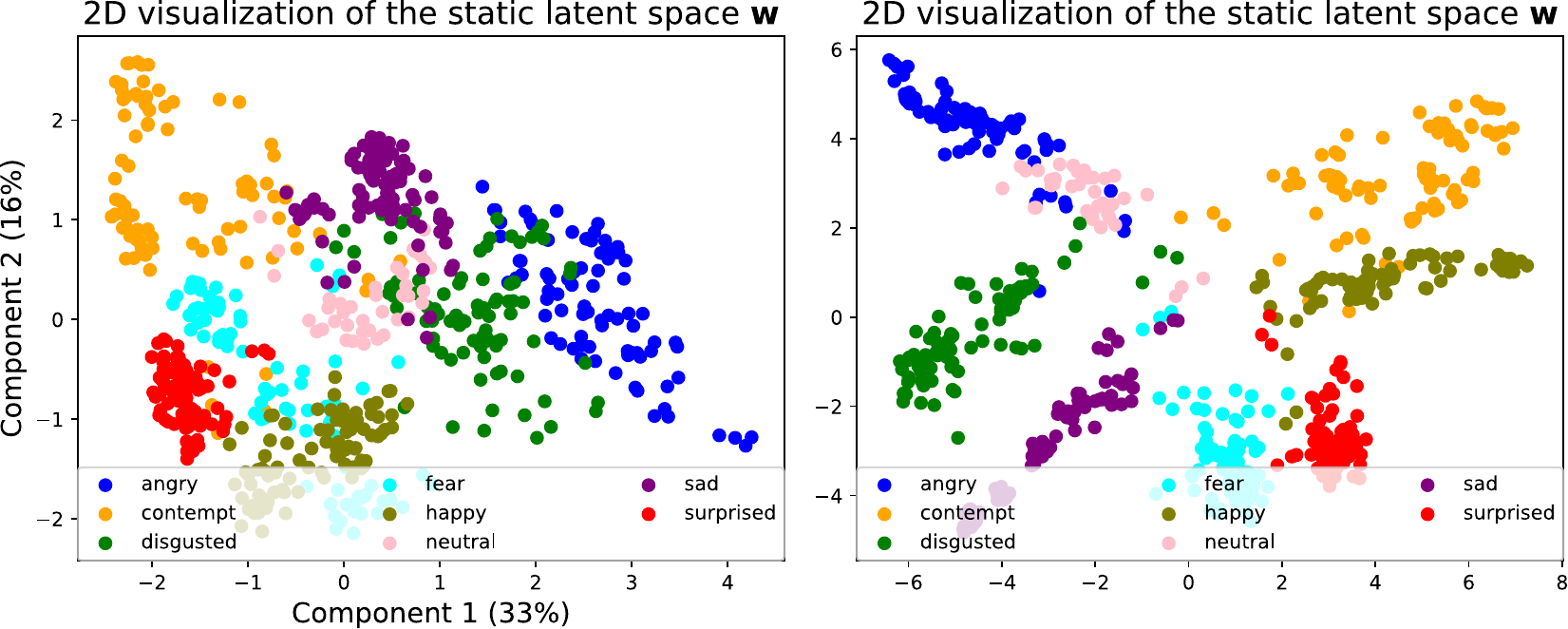}
            \caption{Visualization of the emotion clusters for a single speaker using PCA (left) and ISOMAP (right).}
            \label{fig:viz-emotions}
        \end{subfigure}
     \hfill \vspace{0.1cm}
     \begin{subfigure}[b]{0.48\textwidth}
         \centering
            \includegraphics[width=\textwidth]{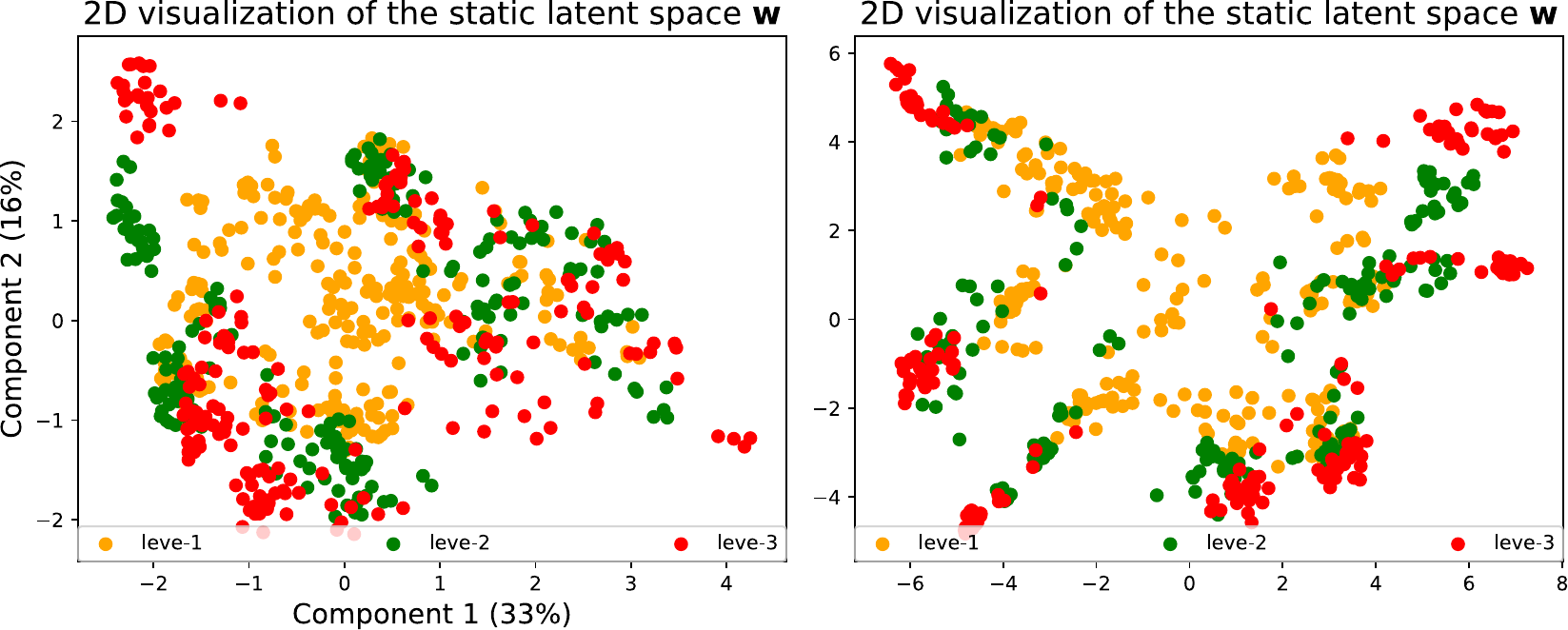}
            \caption{Visualization of the emotional intensity levels for the same speaker as in Figure~\ref{fig:viz-emotions} using PCA (left) and ISOMAP (right).}
            \label{fig:viz-level}
     \end{subfigure}
     \begin{subfigure}[b]{0.48\textwidth}
         \centering
            \includegraphics[width=\textwidth]{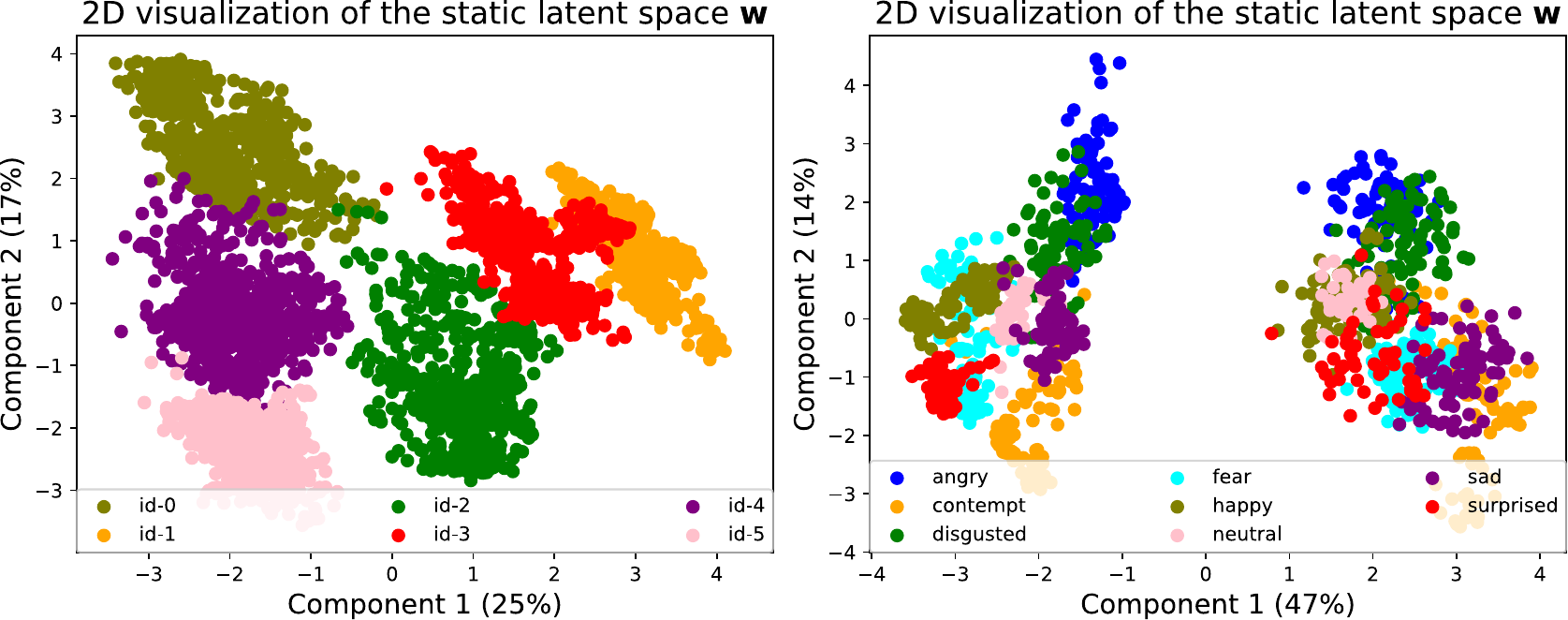}
            \caption{Visualization of the identity clusters for six speakers using PCA (left) and visualization of the emotion clusters for two speakers using PCA (right).}
            \label{fig:viz-identity}
     \end{subfigure}
        \caption{2D visualizations of the static latent space.}
        \label{fig:2D-visualisation}
\end{figure}

2D visualizations of the static latent space of the MDVAE are obtained using dimension reduction methods. Figure~\ref{fig:viz-emotions} shows visualizations obtained with PCA and ISOMAP for one single speaker in the MEAD dataset, and the colors indicate the emotion labels. It can be seen that different emotions form different clusters and the neutral emotion is approximately in the middle. Figure~\ref{fig:viz-level} corresponds to the same visualization but the colors now indicate the emotion intensity levels. It can be seen that for each emotion, the intensity level increases continuously from the middle to the outside of the emotion cluster. Finally, Figure~\ref{fig:viz-identity} shows the identity clusters for six different speakers (left figure) and the emotion clusters for two speakers (right figure), both obtained using PCA. 3D visualizations are available on the companion website, along with other dimension reduction methods.

\bibliography{mybibfile}

\end{document}

%% file: math_commands.tex
\usepackage{amsmath,amsfonts,bm}

\def\eqref#1{equation~(\ref{#1})}
\def\Eqref#1{Equation~(\ref{#1})}
\def\plaineqref#1{(\ref{#1})}

\def\1{\bm{1}}

\DeclareMathAlphabet{\mathsfit}{\encodingdefault}{\sfdefault}{m}{sl}
\SetMathAlphabet{\mathsfit}{bold}{\encodingdefault}{\sfdefault}{bx}{n}













%% file: main.bbl
\begin{thebibliography}{73}
\expandafter\ifx\csname natexlab\endcsname\relax\def\natexlab#1{#1}\fi
\providecommand{\url}[1]{\texttt{#1}}
\providecommand{\href}[2]{#2}
\providecommand{\path}[1]{#1}
\providecommand{\DOIprefix}{doi:}
\providecommand{\ArXivprefix}{arXiv:}
\providecommand{\URLprefix}{URL: }
\providecommand{\Pubmedprefix}{pmid:}
\providecommand{\doi}[1]{\href{http://dx.doi.org/#1}{\path{#1}}}
\providecommand{\Pubmed}[1]{\href{pmid:#1}{\path{#1}}}
\providecommand{\bibinfo}[2]{#2}
\ifx\xfnm\relax \def\xfnm[#1]{\unskip,\space#1}\fi
\bibitem[{Afouras et~al.(2018)Afouras, Chung, Senior, Vinyals \& Zisserman}]{afouras2018deep}
\bibinfo{author}{Afouras, T.}, \bibinfo{author}{Chung, J.~S.}, \bibinfo{author}{Senior, A.}, \bibinfo{author}{Vinyals, O.}, \& \bibinfo{author}{Zisserman, A.} (\bibinfo{year}{2018}).
\newblock \bibinfo{title}{Deep audio-visual speech recognition}.
\newblock {\it \bibinfo{journal}{IEEE transactions on pattern analysis and machine intelligence}\/}, .
\bibitem[{Arnela et~al.(2016)Arnela, Blandin, Dabbaghchian, Guasch, Al{\'\i}as, Pelorson, Van~Hirtum \& Engwall}]{arnela2016influence}
\bibinfo{author}{Arnela, M.}, \bibinfo{author}{Blandin, R.}, \bibinfo{author}{Dabbaghchian, S.}, \bibinfo{author}{Guasch, O.}, \bibinfo{author}{Al{\'\i}as, F.}, \bibinfo{author}{Pelorson, X.}, \bibinfo{author}{Van~Hirtum, A.}, \& \bibinfo{author}{Engwall, O.} (\bibinfo{year}{2016}).
\newblock \bibinfo{title}{Influence of lips on the production of vowels based on finite element simulations and experiments}.
\newblock {\it \bibinfo{journal}{The Journal of the Acoustical Society of America}\/},  {\it \bibinfo{volume}{139}\/}, \bibinfo{pages}{2852--2859}.
\bibitem[{Bachmann et~al.(2022)Bachmann, Mizrahi, Atanov \& Zamir}]{bachmann2022multimae}
\bibinfo{author}{Bachmann, R.}, \bibinfo{author}{Mizrahi, D.}, \bibinfo{author}{Atanov, A.}, \& \bibinfo{author}{Zamir, A.} (\bibinfo{year}{2022}).
\newblock \bibinfo{title}{Multimae: Multi-modal multi-task masked autoencoders}.
\newblock In {\it \bibinfo{booktitle}{European Conference on Computer Vision}\/} (pp. \bibinfo{pages}{348--367}).
\newblock \bibinfo{organization}{Springer}.
\bibitem[{Baltru{\v{s}}aitis et~al.(2018)Baltru{\v{s}}aitis, Ahuja \& Morency}]{baltruvsaitis2018multimodal}
\bibinfo{author}{Baltru{\v{s}}aitis, T.}, \bibinfo{author}{Ahuja, C.}, \& \bibinfo{author}{Morency, L.-P.} (\bibinfo{year}{2018}).
\newblock \bibinfo{title}{Multimodal machine learning: A survey and taxonomy}.
\newblock {\it \bibinfo{journal}{IEEE transactions on pattern analysis and machine intelligence}\/},  {\it \bibinfo{volume}{41}\/}, \bibinfo{pages}{423--443}.
\bibitem[{Baltru{\v{s}}aitis et~al.(2016)Baltru{\v{s}}aitis, Robinson \& Morency}]{baltruvsaitis2016openface}
\bibinfo{author}{Baltru{\v{s}}aitis, T.}, \bibinfo{author}{Robinson, P.}, \& \bibinfo{author}{Morency, L.-P.} (\bibinfo{year}{2016}).
\newblock \bibinfo{title}{Openface: an open source facial behavior analysis toolkit}.
\newblock In {\it \bibinfo{booktitle}{2016 IEEE Winter Conference on Applications of Computer Vision (WACV)}\/} (pp. \bibinfo{pages}{1--10}).
\newblock \bibinfo{organization}{IEEE}.
\bibitem[{Bengio et~al.(2013)Bengio, Courville \& Vincent}]{bengio2013representation}
\bibinfo{author}{Bengio, Y.}, \bibinfo{author}{Courville, A.}, \& \bibinfo{author}{Vincent, P.} (\bibinfo{year}{2013}).
\newblock \bibinfo{title}{Representation learning: A review and new perspectives}.
\newblock {\it \bibinfo{journal}{IEEE transactions on pattern analysis and machine intelligence}\/},  {\it \bibinfo{volume}{35}\/}, \bibinfo{pages}{1798--1828}.
\bibitem[{Berry et~al.(2022)Berry, Lewin \& Brown}]{berry2022correlated}
\bibinfo{author}{Berry, M.}, \bibinfo{author}{Lewin, S.}, \& \bibinfo{author}{Brown, S.} (\bibinfo{year}{2022}).
\newblock \bibinfo{title}{Correlated expression of the body, face, and voice during character portrayal in actors}.
\newblock {\it \bibinfo{journal}{Scientific Reports}\/},  {\it \bibinfo{volume}{12}\/}, \bibinfo{pages}{1--13}.
\bibitem[{Bishop \& Nasrabadi(2006)}]{bishop2006pattern}
\bibinfo{author}{Bishop, C.~M.}, \& \bibinfo{author}{Nasrabadi, N.~M.} (\bibinfo{year}{2006}).
\newblock {\it \bibinfo{title}{Pattern recognition and machine learning}\/} volume~\bibinfo{volume}{4}.
\newblock \bibinfo{publisher}{Springer}.
\bibitem[{Boersma \& Weenink(2021)}]{boersma2021praat}
\bibinfo{author}{Boersma, P.}, \& \bibinfo{author}{Weenink, D.} (\bibinfo{year}{2021}).
\newblock \bibinfo{title}{Praat: doing phonetics by computer [computer program](2011)}.
\newblock {\it \bibinfo{journal}{Version}\/},  {\it \bibinfo{volume}{5}\/}, \bibinfo{pages}{74}.
\bibitem[{Chen et~al.(2018)Chen, Li, Grosse \& Duvenaud}]{chen2018isolating}
\bibinfo{author}{Chen, R.~T.}, \bibinfo{author}{Li, X.}, \bibinfo{author}{Grosse, R.~B.}, \& \bibinfo{author}{Duvenaud, D.~K.} (\bibinfo{year}{2018}).
\newblock \bibinfo{title}{Isolating sources of disentanglement in variational autoencoders}.
\newblock {\it \bibinfo{journal}{Advances in neural information processing systems}\/},  {\it \bibinfo{volume}{31}\/}.
\bibitem[{Chumachenko et~al.(2022)Chumachenko, Iosifidis \& Gabbouj}]{chumachenko2022self}
\bibinfo{author}{Chumachenko, K.}, \bibinfo{author}{Iosifidis, A.}, \& \bibinfo{author}{Gabbouj, M.} (\bibinfo{year}{2022}).
\newblock \bibinfo{title}{Self-attention fusion for audiovisual emotion recognition with incomplete data}.
\newblock In {\it \bibinfo{booktitle}{International Conference on Pattern Recognition (ICPR)}\/} (pp. \bibinfo{pages}{2822--2828}).
\newblock \bibinfo{organization}{IEEE}.
\bibitem[{Courty et~al.(2017)Courty, Flamary, Habrard \& Rakotomamonjy}]{courty2017joint}
\bibinfo{author}{Courty, N.}, \bibinfo{author}{Flamary, R.}, \bibinfo{author}{Habrard, A.}, \& \bibinfo{author}{Rakotomamonjy, A.} (\bibinfo{year}{2017}).
\newblock \bibinfo{title}{Joint distribution optimal transportation for domain adaptation}.
\newblock {\it \bibinfo{journal}{Advances in neural information processing systems}\/},  {\it \bibinfo{volume}{30}\/}.
\bibitem[{Daunhawer et~al.(2021)Daunhawer, Sutter, Chin-Cheong, Palumbo \& Vogt}]{daunhawer2021limitations}
\bibinfo{author}{Daunhawer, I.}, \bibinfo{author}{Sutter, T.~M.}, \bibinfo{author}{Chin-Cheong, K.}, \bibinfo{author}{Palumbo, E.}, \& \bibinfo{author}{Vogt, J.~E.} (\bibinfo{year}{2021}).
\newblock \bibinfo{title}{On the limitations of multimodal vaes}.
\newblock In {\it \bibinfo{booktitle}{International Conference on Learning Representations (ICLR)}\/}.
\bibitem[{Ekman \& Friesen(1978)}]{ekman1978facial}
\bibinfo{author}{Ekman, P.}, \& \bibinfo{author}{Friesen, W.~V.} (\bibinfo{year}{1978}).
\newblock \bibinfo{title}{Facial action coding system}.
\newblock {\it \bibinfo{journal}{Environmental Psychology \& Nonverbal Behavior}\/}, .
\bibitem[{F{\'e}votte et~al.(2009)F{\'e}votte, Bertin \& Durrieu}]{fevotte2009nonnegative}
\bibinfo{author}{F{\'e}votte, C.}, \bibinfo{author}{Bertin, N.}, \& \bibinfo{author}{Durrieu, J.-L.} (\bibinfo{year}{2009}).
\newblock \bibinfo{title}{Nonnegative matrix factorization with the itakura-saito divergence: With application to music analysis}.
\newblock {\it \bibinfo{journal}{Neural computation}\/},  {\it \bibinfo{volume}{21}\/}, \bibinfo{pages}{793--830}.
\bibitem[{Gao \& Shinkareva(2021)}]{gao2021modality}
\bibinfo{author}{Gao, C.}, \& \bibinfo{author}{Shinkareva, S.~V.} (\bibinfo{year}{2021}).
\newblock \bibinfo{title}{Modality-general and modality-specific audiovisual valence processing}.
\newblock {\it \bibinfo{journal}{Cortex}\/},  {\it \bibinfo{volume}{138}\/}, \bibinfo{pages}{127--137}.
\bibitem[{Geiger et~al.(1990)Geiger, Verma \& Pearl}]{geiger1990identifying}
\bibinfo{author}{Geiger, D.}, \bibinfo{author}{Verma, T.}, \& \bibinfo{author}{Pearl, J.} (\bibinfo{year}{1990}).
\newblock \bibinfo{title}{Identifying independence in bayesian networks}.
\newblock {\it \bibinfo{journal}{Networks}\/},  {\it \bibinfo{volume}{20}\/}, \bibinfo{pages}{507--534}.
\bibitem[{Girin et~al.(2021)Girin, Leglaive, Bie, Diard, Hueber \& Alameda-Pineda}]{girin2021dynamical}
\bibinfo{author}{Girin, L.}, \bibinfo{author}{Leglaive, S.}, \bibinfo{author}{Bie, X.}, \bibinfo{author}{Diard, J.}, \bibinfo{author}{Hueber, T.}, \& \bibinfo{author}{Alameda-Pineda, X.} (\bibinfo{year}{2021}).
\newblock \bibinfo{title}{Dynamical variational autoencoders: A comprehensive review}.
\newblock {\it \bibinfo{journal}{Foundations and Trends in Machine Learning}\/},  {\it \bibinfo{volume}{15}\/}, \bibinfo{pages}{1--175}.
\bibitem[{Goodfellow et~al.(2014)Goodfellow, Pouget-Abadie, Mirza, Xu, Warde-Farley, Ozair, Courville \& Bengio}]{goodfellow2014generative}
\bibinfo{author}{Goodfellow, I.}, \bibinfo{author}{Pouget-Abadie, J.}, \bibinfo{author}{Mirza, M.}, \bibinfo{author}{Xu, B.}, \bibinfo{author}{Warde-Farley, D.}, \bibinfo{author}{Ozair, S.}, \bibinfo{author}{Courville, A.}, \& \bibinfo{author}{Bengio, Y.} (\bibinfo{year}{2014}).
\newblock \bibinfo{title}{Generative adversarial nets}.
\newblock {\it \bibinfo{journal}{Advances in neural information processing systems}\/},  {\it \bibinfo{volume}{27}\/}.
\bibitem[{Higgins et~al.(2016)Higgins, Matthey, Pal, Burgess, Glorot, Botvinick, Mohamed \& Lerchner}]{higgins2016beta}
\bibinfo{author}{Higgins, I.}, \bibinfo{author}{Matthey, L.}, \bibinfo{author}{Pal, A.}, \bibinfo{author}{Burgess, C.}, \bibinfo{author}{Glorot, X.}, \bibinfo{author}{Botvinick, M.}, \bibinfo{author}{Mohamed, S.}, \& \bibinfo{author}{Lerchner, A.} (\bibinfo{year}{2016}).
\newblock \bibinfo{title}{beta-vae: Learning basic visual concepts with a constrained variational framework}, .
\bibitem[{Hinton(2002)}]{hinton2002training}
\bibinfo{author}{Hinton, G.~E.} (\bibinfo{year}{2002}).
\newblock \bibinfo{title}{Training products of experts by minimizing contrastive divergence}.
\newblock {\it \bibinfo{journal}{Neural computation}\/},  {\it \bibinfo{volume}{14}\/}, \bibinfo{pages}{1771--1800}.
\bibitem[{Hori et~al.(2019)Hori, Alamri, Wang, Wichern, Hori, Cherian, Marks, Cartillier, Lopes, Das et~al.}]{hori2019end}
\bibinfo{author}{Hori, C.}, \bibinfo{author}{Alamri, H.}, \bibinfo{author}{Wang, J.}, \bibinfo{author}{Wichern, G.}, \bibinfo{author}{Hori, T.}, \bibinfo{author}{Cherian, A.}, \bibinfo{author}{Marks, T.~K.}, \bibinfo{author}{Cartillier, V.}, \bibinfo{author}{Lopes, R.~G.}, \bibinfo{author}{Das, A.} et~al. (\bibinfo{year}{2019}).
\newblock \bibinfo{title}{End-to-end audio visual scene-aware dialog using multimodal attention-based video features}.
\newblock In {\it \bibinfo{booktitle}{IEEE International Conference on Acoustics, Speech and Signal Processing (ICASSP)}\/} (pp. \bibinfo{pages}{2352--2356}).
\newblock \bibinfo{organization}{IEEE}.
\bibitem[{Hou et~al.(2019)Hou, Sun, Shen \& Qiu}]{hou2019improving}
\bibinfo{author}{Hou, X.}, \bibinfo{author}{Sun, K.}, \bibinfo{author}{Shen, L.}, \& \bibinfo{author}{Qiu, G.} (\bibinfo{year}{2019}).
\newblock \bibinfo{title}{Improving variational autoencoder with deep feature consistent and generative adversarial training}.
\newblock {\it \bibinfo{journal}{Neurocomputing}\/},  {\it \bibinfo{volume}{341}\/}, \bibinfo{pages}{183--194}.
\bibitem[{Hsu \& Glass(2018)}]{hsu2018disentangling}
\bibinfo{author}{Hsu, W.-N.}, \& \bibinfo{author}{Glass, J.} (\bibinfo{year}{2018}).
\newblock \bibinfo{title}{Disentangling by partitioning: A representation learning framework for multimodal sensory data}.
\newblock \bibinfo{note}{ArXiv preprint arXiv:1805.11264}.
\bibitem[{Jordan et~al.(1999)Jordan, Ghahramani, Jaakkola \& Saul}]{jordan1999introduction}
\bibinfo{author}{Jordan, M.~I.}, \bibinfo{author}{Ghahramani, Z.}, \bibinfo{author}{Jaakkola, T.~S.}, \& \bibinfo{author}{Saul, L.~K.} (\bibinfo{year}{1999}).
\newblock \bibinfo{title}{An introduction to variational methods for graphical models}.
\newblock {\it \bibinfo{journal}{Machine learning}\/},  {\it \bibinfo{volume}{37}\/}, \bibinfo{pages}{183--233}.
\bibitem[{Kim \& Song(2022)}]{kim2022optimal}
\bibinfo{author}{Kim, D.}, \& \bibinfo{author}{Song, B.~C.} (\bibinfo{year}{2022}).
\newblock \bibinfo{title}{Optimal transport-based identity matching for identity-invariant facial expression recognition}.
\newblock In {\it \bibinfo{booktitle}{Advances in Neural Information Processing Systems}\/}.
\bibitem[{Kim \& Mnih(2018)}]{kim2018disentangling}
\bibinfo{author}{Kim, H.}, \& \bibinfo{author}{Mnih, A.} (\bibinfo{year}{2018}).
\newblock \bibinfo{title}{Disentangling by factorising}.
\newblock In {\it \bibinfo{booktitle}{International Conference on Machine Learning (ICML)}\/} (pp. \bibinfo{pages}{2649--2658}).
\newblock \bibinfo{organization}{PMLR}.
\bibitem[{Kim et~al.(2018)Kim, Salamon, Li \& Bello}]{kim2018crepe}
\bibinfo{author}{Kim, J.~W.}, \bibinfo{author}{Salamon, J.}, \bibinfo{author}{Li, P.}, \& \bibinfo{author}{Bello, J.~P.} (\bibinfo{year}{2018}).
\newblock \bibinfo{title}{Crepe: A convolutional representation for pitch estimation}.
\newblock In {\it \bibinfo{booktitle}{2018 IEEE International Conference on Acoustics, Speech and Signal Processing (ICASSP)}\/} (pp. \bibinfo{pages}{161--165}).
\newblock \bibinfo{organization}{IEEE}.
\bibitem[{Kingma \& Welling(2014)}]{kingma2014auto}
\bibinfo{author}{Kingma, D.}, \& \bibinfo{author}{Welling, M.} (\bibinfo{year}{2014}).
\newblock \bibinfo{title}{Auto-encoding variational bayes.}
\newblock In {\it \bibinfo{booktitle}{International Conference on Learning Representations (ICLR)}\/}.
\bibitem[{Kingma \& Ba(2015)}]{DBLP:journals/corr/KingmaB14}
\bibinfo{author}{Kingma, D.~P.}, \& \bibinfo{author}{Ba, J.} (\bibinfo{year}{2015}).
\newblock \bibinfo{title}{Adam: {A} method for stochastic optimization}.
\newblock In {\it \bibinfo{booktitle}{International Conference on Learning Representations (ICLR)}\/}.
\bibitem[{Klys et~al.(2018)Klys, Snell \& Zemel}]{klys2018learning}
\bibinfo{author}{Klys, J.}, \bibinfo{author}{Snell, J.}, \& \bibinfo{author}{Zemel, R.} (\bibinfo{year}{2018}).
\newblock \bibinfo{title}{Learning latent subspaces in variational autoencoders}.
\newblock {\it \bibinfo{journal}{Advances in neural information processing systems}\/},  {\it \bibinfo{volume}{31}\/}.
\bibitem[{Larsen et~al.(2016)Larsen, S{\o}nderby, Larochelle \& Winther}]{larsen2016autoencoding}
\bibinfo{author}{Larsen, A. B.~L.}, \bibinfo{author}{S{\o}nderby, S.~K.}, \bibinfo{author}{Larochelle, H.}, \& \bibinfo{author}{Winther, O.} (\bibinfo{year}{2016}).
\newblock \bibinfo{title}{Autoencoding beyond pixels using a learned similarity metric}.
\newblock In {\it \bibinfo{booktitle}{International conference on machine learning (ICML)}\/} (pp. \bibinfo{pages}{1558--1566}).
\newblock \bibinfo{organization}{PMLR}.
\bibitem[{Lazarus(1976)}]{lazarus1976multimodal}
\bibinfo{author}{Lazarus, A.~A.} (\bibinfo{year}{1976}).
\newblock \bibinfo{title}{Multimodal therapy}.
\newblock {\it \bibinfo{journal}{Handbook of Psychotherapy Integration}\/},  (p. \bibinfo{pages}{105}).
\bibitem[{Le~Roux et~al.(2019)Le~Roux, Wisdom, Erdogan \& Hershey}]{le2019sdr}
\bibinfo{author}{Le~Roux, J.}, \bibinfo{author}{Wisdom, S.}, \bibinfo{author}{Erdogan, H.}, \& \bibinfo{author}{Hershey, J.~R.} (\bibinfo{year}{2019}).
\newblock \bibinfo{title}{Sdr--half-baked or well done?}
\newblock In {\it \bibinfo{booktitle}{ICASSP 2019-2019 IEEE International Conference on Acoustics, Speech and Signal Processing (ICASSP)}\/} (pp. \bibinfo{pages}{626--630}).
\newblock \bibinfo{organization}{IEEE}.
\bibitem[{Lee \& Pavlovic(2020)}]{lee2020private}
\bibinfo{author}{Lee, M.}, \& \bibinfo{author}{Pavlovic, V.} (\bibinfo{year}{2020}).
\newblock \bibinfo{title}{Private-shared disentangled multimodal vae for learning of hybrid latent representations}.
\newblock \bibinfo{note}{ArXiv preprint arXiv:2012.13024}.
\bibitem[{Li \& Mandt(2018)}]{li2018disentangled}
\bibinfo{author}{Li, Y.}, \& \bibinfo{author}{Mandt, S.} (\bibinfo{year}{2018}).
\newblock \bibinfo{title}{Disentangled sequential autoencoder}.
\newblock \bibinfo{note}{ArXiv preprint arXiv:1803.02991}.
\bibitem[{Livingstone \& Russo(2018)}]{livingstone2018ryerson}
\bibinfo{author}{Livingstone, S.~R.}, \& \bibinfo{author}{Russo, F.~A.} (\bibinfo{year}{2018}).
\newblock \bibinfo{title}{The ryerson audio-visual database of emotional speech and song (ravdess): A dynamic, multimodal set of facial and vocal expressions in north american english}.
\newblock {\it \bibinfo{journal}{PloS one}\/},  {\it \bibinfo{volume}{13}\/}, \bibinfo{pages}{e0196391}.
\bibitem[{Lo et~al.(2019)Lo, Fu, Huang, Wang, Yamagishi, Tsao \& Wang}]{lo2019mosnet}
\bibinfo{author}{Lo, C.-C.}, \bibinfo{author}{Fu, S.-W.}, \bibinfo{author}{Huang, W.-C.}, \bibinfo{author}{Wang, X.}, \bibinfo{author}{Yamagishi, J.}, \bibinfo{author}{Tsao, Y.}, \& \bibinfo{author}{Wang, H.-M.} (\bibinfo{year}{2019}).
\newblock \bibinfo{title}{Mosnet: Deep learning based objective assessment for voice conversion}.
\newblock \bibinfo{note}{ArXiv preprint arXiv:1904.08352}.
\bibitem[{Locatello et~al.(2019)Locatello, Bauer, Lucic, Raetsch, Gelly, Sch{\"o}lkopf \& Bachem}]{locatello2019challenging}
\bibinfo{author}{Locatello, F.}, \bibinfo{author}{Bauer, S.}, \bibinfo{author}{Lucic, M.}, \bibinfo{author}{Raetsch, G.}, \bibinfo{author}{Gelly, S.}, \bibinfo{author}{Sch{\"o}lkopf, B.}, \& \bibinfo{author}{Bachem, O.} (\bibinfo{year}{2019}).
\newblock \bibinfo{title}{Challenging common assumptions in the unsupervised learning of disentangled representations}.
\newblock In {\it \bibinfo{booktitle}{International conference on machine learning (ICML)}\/} (pp. \bibinfo{pages}{4114--4124}).
\newblock \bibinfo{organization}{PMLR}.
\bibitem[{Locatello et~al.(2020)Locatello, Poole, R{\"a}tsch, Sch{\"o}lkopf, Bachem \& Tschannen}]{locatello2020weakly}
\bibinfo{author}{Locatello, F.}, \bibinfo{author}{Poole, B.}, \bibinfo{author}{R{\"a}tsch, G.}, \bibinfo{author}{Sch{\"o}lkopf, B.}, \bibinfo{author}{Bachem, O.}, \& \bibinfo{author}{Tschannen, M.} (\bibinfo{year}{2020}).
\newblock \bibinfo{title}{Weakly-supervised disentanglement without compromises}.
\newblock In {\it \bibinfo{booktitle}{International Conference on Machine Learning (ICML)}\/} (pp. \bibinfo{pages}{6348--6359}).
\newblock \bibinfo{organization}{PMLR}.
\bibitem[{Mollahosseini et~al.(2017)Mollahosseini, Hasani \& Mahoor}]{mollahosseini2017affectnet}
\bibinfo{author}{Mollahosseini, A.}, \bibinfo{author}{Hasani, B.}, \& \bibinfo{author}{Mahoor, M.~H.} (\bibinfo{year}{2017}).
\newblock \bibinfo{title}{Affectnet: A database for facial expression, valence, and arousal computing in the wild}.
\newblock {\it \bibinfo{journal}{IEEE Transactions on Affective Computing}\/},  {\it \bibinfo{volume}{10}\/}, \bibinfo{pages}{18--31}.
\bibitem[{Muhammod et~al.(2019)Muhammod, Ahmed, Md~Farid, Shatabda, Sharma \& Dehzangi}]{muhammod2019pyfeat}
\bibinfo{author}{Muhammod, R.}, \bibinfo{author}{Ahmed, S.}, \bibinfo{author}{Md~Farid, D.}, \bibinfo{author}{Shatabda, S.}, \bibinfo{author}{Sharma, A.}, \& \bibinfo{author}{Dehzangi, A.} (\bibinfo{year}{2019}).
\newblock \bibinfo{title}{Pyfeat: a python-based effective feature generation tool for dna, rna and protein sequences}.
\newblock {\it \bibinfo{journal}{Bioinformatics}\/},  {\it \bibinfo{volume}{35}\/}, \bibinfo{pages}{3831--3833}.
\bibitem[{Neal \& Hinton(1998)}]{neal1998view}
\bibinfo{author}{Neal, R.~M.}, \& \bibinfo{author}{Hinton, G.~E.} (\bibinfo{year}{1998}).
\newblock \bibinfo{title}{A view of the em algorithm that justifies incremental, sparse, and other variants}.
\newblock In {\it \bibinfo{booktitle}{Learning in graphical models}\/} (pp. \bibinfo{pages}{355--368}).
\newblock \bibinfo{publisher}{Springer}.
\bibitem[{Noroozi et~al.(2017)Noroozi, Marjanovic, Njegus, Escalera \& Anbarjafari}]{noroozi2017audio}
\bibinfo{author}{Noroozi, F.}, \bibinfo{author}{Marjanovic, M.}, \bibinfo{author}{Njegus, A.}, \bibinfo{author}{Escalera, S.}, \& \bibinfo{author}{Anbarjafari, G.} (\bibinfo{year}{2017}).
\newblock \bibinfo{title}{Audio-visual emotion recognition in video clips}.
\newblock {\it \bibinfo{journal}{IEEE Transactions on Affective Computing}\/},  {\it \bibinfo{volume}{10}\/}, \bibinfo{pages}{60--75}.
\bibitem[{Panayotov et~al.(2015)Panayotov, Chen, Povey \& Khudanpur}]{panayotov2015librispeech}
\bibinfo{author}{Panayotov, V.}, \bibinfo{author}{Chen, G.}, \bibinfo{author}{Povey, D.}, \& \bibinfo{author}{Khudanpur, S.} (\bibinfo{year}{2015}).
\newblock \bibinfo{title}{Librispeech: an asr corpus based on public domain audio books}.
\newblock In {\it \bibinfo{booktitle}{IEEE international conference on acoustics, speech and signal processing (ICASSP)}\/} (pp. \bibinfo{pages}{5206--5210}).
\newblock \bibinfo{organization}{IEEE}.
\bibitem[{Pepino et~al.(2021)Pepino, Riera \& Ferrer}]{pepino2021emotion}
\bibinfo{author}{Pepino, L.}, \bibinfo{author}{Riera, P.}, \& \bibinfo{author}{Ferrer, L.} (\bibinfo{year}{2021}).
\newblock \bibinfo{title}{Emotion recognition from speech using wav2vec 2.0 embeddings}.
\newblock {\it \bibinfo{journal}{Interspeech}\/},  (pp. \bibinfo{pages}{3400--3404}).
\bibitem[{Petridis et~al.(2018)Petridis, Stafylakis, Ma, Cai, Tzimiropoulos \& Pantic}]{petridis2018end}
\bibinfo{author}{Petridis, S.}, \bibinfo{author}{Stafylakis, T.}, \bibinfo{author}{Ma, P.}, \bibinfo{author}{Cai, F.}, \bibinfo{author}{Tzimiropoulos, G.}, \& \bibinfo{author}{Pantic, M.} (\bibinfo{year}{2018}).
\newblock \bibinfo{title}{End-to-end audiovisual speech recognition}.
\newblock In {\it \bibinfo{booktitle}{IEEE international conference on acoustics, speech and signal processing (ICASSP)}\/} (pp. \bibinfo{pages}{6548--6552}).
\newblock \bibinfo{organization}{IEEE}.
\bibitem[{Pham et~al.(2021)Pham, Vu \& Tran}]{pham2021facial}
\bibinfo{author}{Pham, L.}, \bibinfo{author}{Vu, T.~H.}, \& \bibinfo{author}{Tran, T.~A.} (\bibinfo{year}{2021}).
\newblock \bibinfo{title}{Facial expression recognition using residual masking network}.
\newblock In {\it \bibinfo{booktitle}{International Conference on Pattern Recognition (ICPR)}\/} (pp. \bibinfo{pages}{4513--4519}).
\newblock \bibinfo{organization}{IEEE}.
\bibitem[{Pihlgren et~al.(2020)Pihlgren, Sandin \& Liwicki}]{pihlgren2020improving}
\bibinfo{author}{Pihlgren, G.~G.}, \bibinfo{author}{Sandin, F.}, \& \bibinfo{author}{Liwicki, M.} (\bibinfo{year}{2020}).
\newblock \bibinfo{title}{Improving image autoencoder embeddings with perceptual loss}.
\newblock In {\it \bibinfo{booktitle}{2020 International Joint Conference on Neural Networks (IJCNN)}\/} (pp. \bibinfo{pages}{1--7}).
\newblock \bibinfo{organization}{IEEE}.
\bibitem[{Ramachandram \& Taylor(2017)}]{ramachandram2017deep}
\bibinfo{author}{Ramachandram, D.}, \& \bibinfo{author}{Taylor, G.~W.} (\bibinfo{year}{2017}).
\newblock \bibinfo{title}{Deep multimodal learning: A survey on recent advances and trends}.
\newblock {\it \bibinfo{journal}{IEEE signal processing magazine}\/},  {\it \bibinfo{volume}{34}\/}, \bibinfo{pages}{96--108}.
\bibitem[{Razavi et~al.(2019)Razavi, Van~den Oord \& Vinyals}]{razavi2019generating}
\bibinfo{author}{Razavi, A.}, \bibinfo{author}{Van~den Oord, A.}, \& \bibinfo{author}{Vinyals, O.} (\bibinfo{year}{2019}).
\newblock \bibinfo{title}{Generating diverse high-fidelity images with vq-vae-2}.
\newblock {\it \bibinfo{journal}{Advances in neural information processing systems}\/},  {\it \bibinfo{volume}{32}\/}.
\bibitem[{Rezende et~al.(2014)Rezende, Mohamed \& Wierstra}]{rezende2014stochastic}
\bibinfo{author}{Rezende, D.~J.}, \bibinfo{author}{Mohamed, S.}, \& \bibinfo{author}{Wierstra, D.} (\bibinfo{year}{2014}).
\newblock \bibinfo{title}{Stochastic backpropagation and approximate inference in deep generative models}.
\newblock In {\it \bibinfo{booktitle}{International conference on machine learning (ICML)}\/} (pp. \bibinfo{pages}{1278--1286}).
\newblock \bibinfo{organization}{PMLR}.
\bibitem[{Rix et~al.(2001)Rix, Beerends, Hollier \& Hekstra}]{rix2001perceptual}
\bibinfo{author}{Rix, A.~W.}, \bibinfo{author}{Beerends, J.~G.}, \bibinfo{author}{Hollier, M.~P.}, \& \bibinfo{author}{Hekstra, A.~P.} (\bibinfo{year}{2001}).
\newblock \bibinfo{title}{Perceptual evaluation of speech quality (pesq)-a new method for speech quality assessment of telephone networks and codecs}.
\newblock In {\it \bibinfo{booktitle}{IEEE international conference on acoustics, speech, and signal processing. Proceedings (Cat. No. 01CH37221)}\/} (pp. \bibinfo{pages}{749--752}).
\newblock \bibinfo{organization}{IEEE} volume~\bibinfo{volume}{2}.
\bibitem[{Roth et~al.(2020)Roth, Chaudhuri, Klejch, Marvin, Gallagher, Kaver, Ramaswamy, Stopczynski, Schmid, Xi et~al.}]{roth2020ava}
\bibinfo{author}{Roth, J.}, \bibinfo{author}{Chaudhuri, S.}, \bibinfo{author}{Klejch, O.}, \bibinfo{author}{Marvin, R.}, \bibinfo{author}{Gallagher, A.}, \bibinfo{author}{Kaver, L.}, \bibinfo{author}{Ramaswamy, S.}, \bibinfo{author}{Stopczynski, A.}, \bibinfo{author}{Schmid, C.}, \bibinfo{author}{Xi, Z.} et~al. (\bibinfo{year}{2020}).
\newblock \bibinfo{title}{Ava active speaker: An audio-visual dataset for active speaker detection}.
\newblock In {\it \bibinfo{booktitle}{IEEE International Conference on Acoustics, Speech and Signal Processing (ICASSP)}\/} (pp. \bibinfo{pages}{4492--4496}).
\newblock \bibinfo{organization}{IEEE}.
\bibitem[{Sadok et~al.(2023)Sadok, Leglaive, Girin, Alameda-Pineda \& S{\'e}guier}]{sadok2023learning}
\bibinfo{author}{Sadok, S.}, \bibinfo{author}{Leglaive, S.}, \bibinfo{author}{Girin, L.}, \bibinfo{author}{Alameda-Pineda, X.}, \& \bibinfo{author}{S{\'e}guier, R.} (\bibinfo{year}{2023}).
\newblock \bibinfo{title}{Learning and controlling the source-filter representation of speech with a variational autoencoder}.
\newblock {\it \bibinfo{journal}{Speech Communication}\/},  {\it \bibinfo{volume}{148}\/}, \bibinfo{pages}{53--65}.
\bibitem[{Schneider et~al.(2019)Schneider, Baevski, Collobert \& Auli}]{schneider2019wav2vec}
\bibinfo{author}{Schneider, S.}, \bibinfo{author}{Baevski, A.}, \bibinfo{author}{Collobert, R.}, \& \bibinfo{author}{Auli, M.} (\bibinfo{year}{2019}).
\newblock \bibinfo{title}{wav2vec: Unsupervised pre-training for speech recognition}.
\newblock {\it \bibinfo{journal}{Interspeech}\/},  (pp. \bibinfo{pages}{3465--3469}).
\bibitem[{Schoneveld et~al.(2021)Schoneveld, Othmani \& Abdelkawy}]{schoneveld2021leveraging}
\bibinfo{author}{Schoneveld, L.}, \bibinfo{author}{Othmani, A.}, \& \bibinfo{author}{Abdelkawy, H.} (\bibinfo{year}{2021}).
\newblock \bibinfo{title}{Leveraging recent advances in deep learning for audio-visual emotion recognition}.
\newblock {\it \bibinfo{journal}{Pattern Recognition Letters}\/},  {\it \bibinfo{volume}{146}\/}, \bibinfo{pages}{1--7}.
\bibitem[{Shi et~al.(2019)Shi, Paige, Torr et~al.}]{shi2019variational}
\bibinfo{author}{Shi, Y.}, \bibinfo{author}{Paige, B.}, \bibinfo{author}{Torr, P.} et~al. (\bibinfo{year}{2019}).
\newblock \bibinfo{title}{Variational mixture-of-experts autoencoders for multi-modal deep generative models}.
\newblock {\it \bibinfo{journal}{Advances in Neural Information Processing Systems}\/},  {\it \bibinfo{volume}{32}\/}.
\bibitem[{Sutter et~al.(2020)Sutter, Daunhawer \& Vogt}]{sutter2020multimodal}
\bibinfo{author}{Sutter, T.}, \bibinfo{author}{Daunhawer, I.}, \& \bibinfo{author}{Vogt, J.} (\bibinfo{year}{2020}).
\newblock \bibinfo{title}{Multimodal generative learning utilizing jensen-shannon-divergence}.
\newblock {\it \bibinfo{journal}{Advances in Neural Information Processing Systems}\/},  {\it \bibinfo{volume}{33}\/}, \bibinfo{pages}{6100--6110}.
\bibitem[{Sutter et~al.(2021)Sutter, Daunhawer \& Vogt}]{sutter2021generalized}
\bibinfo{author}{Sutter, T.~M.}, \bibinfo{author}{Daunhawer, I.}, \& \bibinfo{author}{Vogt, J.~E.} (\bibinfo{year}{2021}).
\newblock \bibinfo{title}{Generalized multimodal elbo}.
\newblock In {\it \bibinfo{booktitle}{International Conference on Learning Representations (ICLR)}\/}.
\bibitem[{Suzuki \& Matsuo(2022)}]{suzuki2022survey}
\bibinfo{author}{Suzuki, M.}, \& \bibinfo{author}{Matsuo, Y.} (\bibinfo{year}{2022}).
\newblock \bibinfo{title}{A survey of multimodal deep generative models}.
\newblock {\it \bibinfo{journal}{Advanced Robotics}\/},  {\it \bibinfo{volume}{36}\/}, \bibinfo{pages}{261--278}.
\bibitem[{Suzuki et~al.(2016)Suzuki, Nakayama \& Matsuo}]{suzuki2016joint}
\bibinfo{author}{Suzuki, M.}, \bibinfo{author}{Nakayama, K.}, \& \bibinfo{author}{Matsuo, Y.} (\bibinfo{year}{2016}).
\newblock \bibinfo{title}{Joint multimodal learning with deep generative models}.
\newblock \bibinfo{note}{ArXiv preprint arXiv:1611.01891}.
\bibitem[{Taal et~al.(2010)Taal, Hendriks, Heusdens \& Jensen}]{taal2010short}
\bibinfo{author}{Taal, C.~H.}, \bibinfo{author}{Hendriks, R.~C.}, \bibinfo{author}{Heusdens, R.}, \& \bibinfo{author}{Jensen, J.} (\bibinfo{year}{2010}).
\newblock \bibinfo{title}{A short-time objective intelligibility measure for time-frequency weighted noisy speech}.
\newblock In {\it \bibinfo{booktitle}{IEEE international conference on acoustics, speech and signal processing}\/} (pp. \bibinfo{pages}{4214--4217}).
\newblock \bibinfo{organization}{IEEE}.
\bibitem[{Tsai et~al.(2019)Tsai, Bai, Liang, Kolter, Morency \& Salakhutdinov}]{tsai2019multimodal}
\bibinfo{author}{Tsai, Y.-H.~H.}, \bibinfo{author}{Bai, S.}, \bibinfo{author}{Liang, P.~P.}, \bibinfo{author}{Kolter, J.~Z.}, \bibinfo{author}{Morency, L.-P.}, \& \bibinfo{author}{Salakhutdinov, R.} (\bibinfo{year}{2019}).
\newblock \bibinfo{title}{Multimodal transformer for unaligned multimodal language sequences}.
\newblock In {\it \bibinfo{booktitle}{Proceedings of the conference. Association for Computational Linguistics. Meeting}\/} (p. \bibinfo{pages}{6558}).
\newblock \bibinfo{organization}{NIH Public Access} volume \bibinfo{volume}{2019}.
\bibitem[{Vahdat \& Kautz(2020)}]{vahdat2020nvae}
\bibinfo{author}{Vahdat, A.}, \& \bibinfo{author}{Kautz, J.} (\bibinfo{year}{2020}).
\newblock \bibinfo{title}{Nvae: A deep hierarchical variational autoencoder}.
\newblock {\it \bibinfo{journal}{Advances in Neural Information Processing Systems}\/},  {\it \bibinfo{volume}{33}\/}, \bibinfo{pages}{19667--19679}.
\bibitem[{Van Den~Oord et~al.(2017)Van Den~Oord, Vinyals et~al.}]{van2017neural}
\bibinfo{author}{Van Den~Oord, A.}, \bibinfo{author}{Vinyals, O.} et~al. (\bibinfo{year}{2017}).
\newblock \bibinfo{title}{Neural discrete representation learning}.
\newblock {\it \bibinfo{journal}{Advances in neural information processing systems}\/},  {\it \bibinfo{volume}{30}\/}.
\bibitem[{Van~Steenkiste et~al.(2019)Van~Steenkiste, Locatello, Schmidhuber \& Bachem}]{van2019disentangled}
\bibinfo{author}{Van~Steenkiste, S.}, \bibinfo{author}{Locatello, F.}, \bibinfo{author}{Schmidhuber, J.}, \& \bibinfo{author}{Bachem, O.} (\bibinfo{year}{2019}).
\newblock \bibinfo{title}{Are disentangled representations helpful for abstract visual reasoning?}
\newblock {\it \bibinfo{journal}{Advances in Neural Information Processing Systems}\/},  {\it \bibinfo{volume}{32}\/}.
\bibitem[{Wang et~al.(2020)Wang, Wu, Song, Yang, Wu, Qian, He, Qiao \& Loy}]{wang2020mead}
\bibinfo{author}{Wang, K.}, \bibinfo{author}{Wu, Q.}, \bibinfo{author}{Song, L.}, \bibinfo{author}{Yang, Z.}, \bibinfo{author}{Wu, W.}, \bibinfo{author}{Qian, C.}, \bibinfo{author}{He, R.}, \bibinfo{author}{Qiao, Y.}, \& \bibinfo{author}{Loy, C.~C.} (\bibinfo{year}{2020}).
\newblock \bibinfo{title}{Mead: A large-scale audio-visual dataset for emotional talking-face generation}.
\newblock In {\it \bibinfo{booktitle}{European Conference on Computer Vision (ECCV)}\/} (pp. \bibinfo{pages}{700--717}).
\newblock \bibinfo{organization}{Springer}.
\bibitem[{Wang et~al.(2004)Wang, Bovik, Sheikh \& Simoncelli}]{wang2004image}
\bibinfo{author}{Wang, Z.}, \bibinfo{author}{Bovik, A.~C.}, \bibinfo{author}{Sheikh, H.~R.}, \& \bibinfo{author}{Simoncelli, E.~P.} (\bibinfo{year}{2004}).
\newblock \bibinfo{title}{Image quality assessment: from error visibility to structural similarity}.
\newblock {\it \bibinfo{journal}{IEEE transactions on image processing}\/},  {\it \bibinfo{volume}{13}\/}, \bibinfo{pages}{600--612}.
\bibitem[{Wei et~al.(2018)Wei, Li, Sun \& Chen}]{wei2018unsupervised}
\bibinfo{author}{Wei, X.}, \bibinfo{author}{Li, H.}, \bibinfo{author}{Sun, J.}, \& \bibinfo{author}{Chen, L.} (\bibinfo{year}{2018}).
\newblock \bibinfo{title}{Unsupervised domain adaptation with regularized optimal transport for multimodal 2d+ 3d facial expression recognition}.
\newblock In {\it \bibinfo{booktitle}{IEEE International Conference on Automatic Face \& Gesture Recognition (FG 2018)}\/} (pp. \bibinfo{pages}{31--37}).
\newblock \bibinfo{organization}{IEEE}.
\bibitem[{Wu et~al.(2014)Wu, Lin \& Wei}]{wu2014survey}
\bibinfo{author}{Wu, C.-H.}, \bibinfo{author}{Lin, J.-C.}, \& \bibinfo{author}{Wei, W.-L.} (\bibinfo{year}{2014}).
\newblock \bibinfo{title}{Survey on audiovisual emotion recognition: databases, features, and data fusion strategies}.
\newblock {\it \bibinfo{journal}{APSIPA transactions on signal and information processing}\/},  {\it \bibinfo{volume}{3}\/}, \bibinfo{pages}{e12}.
\bibitem[{Wu \& Goodman(2018)}]{wu2018multimodal}
\bibinfo{author}{Wu, M.}, \& \bibinfo{author}{Goodman, N.} (\bibinfo{year}{2018}).
\newblock \bibinfo{title}{Multimodal generative models for scalable weakly-supervised learning}.
\newblock {\it \bibinfo{journal}{Advances in Neural Information Processing Systems}\/},  {\it \bibinfo{volume}{31}\/}.
\bibitem[{Zhao et~al.(2021)Zhao, Liu \& Zhou}]{zhao2021robust}
\bibinfo{author}{Zhao, Z.}, \bibinfo{author}{Liu, Q.}, \& \bibinfo{author}{Zhou, F.} (\bibinfo{year}{2021}).
\newblock \bibinfo{title}{Robust lightweight facial expression recognition network with label distribution training}.
\newblock In {\it \bibinfo{booktitle}{Conference on artificial intelligence (AAAI)}\/} (pp. \bibinfo{pages}{3510--3519}).
\newblock volume~\bibinfo{volume}{35}.

\end{thebibliography}
